\documentclass[11pt]{article}
\usepackage{lmodern}
\usepackage[T1]{fontenc}
\usepackage[utf8]{inputenc}
\usepackage[margin=1in]{geometry}
\usepackage{amsmath,amssymb,amsthm}
\usepackage{graphicx}
\usepackage{tabularx}
\usepackage[hidelinks]{hyperref}
\usepackage{microtype}
\newcommand{\parttitle}[1]{\bigskip\noindent{\large\bfseries #1}\par\setcounter{section}{0}\medskip}
\title{Sparse departures from independence in two-way tables: a heteroscedasticity profile and detection boundary, an adaptive higher-criticism gate, and an assumption-lean exact anchor}
\author{William J. Dwyer, MD, MPH, FAAP\\ Department of Mathematics and Statistics,\\ University of Massachusetts Lowell, Lowell, MA, USA\\ \texttt{wjdwyer@trialdesign.com} \\ ORCID 0009-0004-0855-7222}
\date{}
\begin{document}
\maketitle

\begin{abstract}
Two-way contingency tables are tested for independence throughout applied statistics — in genomics, network and text co-occurrence, pharmacovigilance, ecology, and survey cross-tabulation — and the routine test reads asymptotic Gaussian tail probabilities off the table cell by cell. On the tables people actually analyze this is badly miscalibrated: across a corpus of 5,543 real public two-way tables, two-thirds have counts small enough or margins heterogeneous enough that the asymptotic maximum-cell independence scan false-positives at a mean 33\% against a 0.05 target, while an exact margin-conditional anchor holds at about 1\%. The failure is sharpest when the departure is sparse — independence across the bulk of the table with the association concentrated in a few cells — where the omnibus chi-square is inefficient and the sharp object is the detection boundary, the weakest signal any test can find. This paper answers, in one place and in three parts, where in a table a sparse signal can be seen, which combiner attains that limit, and how to calibrate when counts are small. Part I specializes the sparse-signal detection theory of Ingster (1997), Donoho and Jin (2004), and Chhor, Mukherjee and Sen (2024) to the independence test: the two-way table is a heteroscedastic Gaussian sequence whose variance profile is the expected-count table, fixed by the two margins through the product-CV identity $CVe^{2} = (1 + CVr^{2})(1 + CVc^{2}) - 1$, so marginal heterogeneity is exactly the heteroscedasticity profile; a variance-stabilizing root transform gives the exact standardized amplitude $2 \sqrt{m}(\sqrt{1 + a} - 1)$ and a closed-form separation radius, valid under an explicit growing-count condition and provably failing at fixed small counts. Part II casts independence as goodness of fit of the log-linear no-interaction model and shows that higher criticism, with its closed-form Jaeschke-Eicker null, attains that boundary adaptively over the unknown sparsity in the growing-count regime, while being honest that in the very sparse strong regime the maximum-cell rule dominates and that the exact-conditional optimality is conjectural. Part III measures the small-count calibration failure (actual size 0.3 to 0.6, even under uniform margins, so the driver is the per-cell tail, not heterogeneity), removes it with the exact margin-conditional reference laws (measured size 0.003 to 0.009), and gives a routing rule keyed to the margin profile that sends large-count tables to the asymptotic gate and small-count tables to the exact anchor; two-thirds of the corpus route to the exact anchor. Every claim is confirmed numerically from openly deposited, deterministically seeded code; full proofs are in the supplementary derivations companion (D02).
\end{abstract}


\section*{Structure of the paper}

The three parts are a single arc and share one reproducibility package and one derivations companion (D02, supplied as supplementary material). Part I establishes the object and its information limit: the heteroscedasticity profile, the exact signal map, and the closed-form detection boundary, with the exact scope condition that marks where the boundary holds. Part II supplies the adaptive test that attains the boundary without knowing the sparsity, and is candid about where it is instead dominated by the maximum-cell rule. Part III supplies the finite-sample calibration that both preceding parts require in the low-count corner, together with the routing rule and a reporting apparatus, and places the whole apparatus inside the T\_root reporting standard as its sparse-alternative branch. Each part keeps its own section, figure, table, equation, and theorem numbering; cross-references between parts are given by part number. Throughout, $N = RC$ is the number of cells and n the grand total.

\parttitle{Part I. The heteroscedasticity profile and the detection boundary}

\section{Introduction}

Two-way contingency tables are among the most common objects in applied statistics, and testing them for independence is a routine first step: in genomics (a cell-cluster by marker-gene co-occurrence table, a genome-wide table of variant by phenotype counts), in network and text analysis (a degree-structured co-occurrence table), in pharmacovigilance (a drug by adverse-event enrichment table), in ecology (a species co-occurrence table), and in ordinary survey cross-tabulation. In many of these the interesting departure from independence is sparse: the table is independent across the bulk of its cells and the association is concentrated in a few of them — a rare variant enriched in one phenotype, a marker gene specific to one cluster, a single over-represented drug-event pair. Two facts about this setting motivate the paper. First, for a sparse departure the omnibus chi-square is inefficient, so the sharp question is not which test to use but what the weakest detectable signal is — the detection boundary. Second, the calibration the field routinely applies to the per-cell scan, asymptotic Gaussian tail probabilities, is untrustworthy on exactly the tables that carry these signals: across a corpus of 5,543 real public two-way tables (Part III), two-thirds have counts too small or margins too heterogeneous for the asymptotic scan, on which it false-positives at a mean 33\% against a 0.05 target while an exact margin-conditional anchor holds at about 1\%. This paper treats both facts together — it locates the detection boundary for sparse independence departures, gives an adaptive test that attains it, and supplies the exact small-count calibration the boundary analysis and the field both need.

Pearson's chi-square and the likelihood-ratio $G^{2}$ are omnibus tests of independence: they accumulate evidence across all (R-1)(C-1) interaction degrees of freedom and are efficient against a dense departure in which most cells deviate a little. A different and common situation is a sparse departure: independence holds across the bulk of the table and the association lives in a handful of cells, an interaction pocket, a contaminated stratum, a rare-by-rare cell that is over-represented. For sparse alternatives the omnibus statistic is inefficient, and the sharp question is not "which test" but "what is the weakest signal any test can detect," the detection boundary.

Detection boundaries for sparse signals are understood in detail in the Gaussian sequence model, from Ingster (1997) and Donoho and Jin (2004) through sparse binary regression (Mukherjee, Pillai and Lin, 2015), the beta-model on sparse graphs (Mukherjee, Mukherjee and Sen, 2018), and heteroscedastic sequence models where the noise level varies across coordinates (Chhor, Mukherjee and Sen, 2024). This part specializes that theory to the classical test of independence. Our claim is not a new detection phenomenon but a transport with an exact, margin-explicit dictionary: the two-way table is a heteroscedastic sequence model whose heteroscedasticity profile is not estimated but is the closed-form outer product of the margins.

Section 2 states the profile identity and its population-variance convention. Section 3 reduces the sparse independence test to a heteroscedastic Gaussian sequence and gives the exact signal map, with an honest account of the residual dependence. Section 4 states the separation radius, together with the moderate-deviation growth condition it requires and the small-count normal-tail reason it fails without it. Section 5 reports the numerical confirmations, including the demonstration that the failure at small counts is a tail effect, not a heterogeneity effect. Section 6 places the result beside its companions: the adaptive higher-criticism gate (Part II) and the exact margin-conditional anchor (Part III). Full proofs are in the derivations companion D02. Throughout, $N = RC$ is the number of cells and n is the grand total, and the asymptotic regime is N to infinity; note that this N differs from the T\_root papers (Dwyer 2026a, 2026b), where N denotes the sample size.

\section{The expected-count profile and the product-CV identity}

Let the table have row-margin probabilities $r = (r_{1}, ..., r_{R})$, column-margin $c = (c_{1}, ..., c_{C})$, and, under independence, cell probability $p_{ij} = r_{i} c_{j}$ and expected count $m_{ij} = n r_{i} c_{j}$ for a sample of size n. Call ${m_{ij}}$ the expected-count profile; its heterogeneity $CVe = sd(m)/mean(m)$ over the $RC$ cells is the quantity a reporting standard uses to judge whether a table is regular or heterogeneous (Dwyer 2026a, 2026b).

All coefficients of variation here use the population standard deviation (divisors $R, C, RC$; ddof = 0), consistent with treating the margins and the $RC$ cells as finite populations. With the sample convention the identity below acquires (R-1)/R and (C-1)/C factors.

\textbf{Proposition 1 (profile identity).} For the product-margin profile $m_{ij} = n r_{i} c_{j}$,

\begin{equation*}
CVe^{2} = (1 + CVr^{2})(1 + CVc^{2}) - 1,
\end{equation*}

where $CVr, CVc$ are the population coefficients of variation of the row and column margins.

Proposition 1 is the classical multiplicativity of $1 + CV^{2}$ for a product of independent factors (Goodman 1960), specialized to the outer-product profile; the proof is in D02.1. The identity itself is not new. What it buys is interpretive: the cell-count coefficient of variation equals the heteroscedasticity profile of the sequence model of Section 3, and it factors through the two margins. Three readings follow. The joint heterogeneity factors as $1 + CVe^{2} = (1 + CVr^{2})(1 + CVc^{2})$, never as a sum; a table can be severely heterogeneous through one margin alone. $CVe = 0$ if and only if both margins are uniform, the unique homoscedastic point. And there is no free heteroscedasticity beyond the two margins, which is what makes the reduction of Section 3 fully explicit and lets the calibration axis $CVe$ double as the detection-problem profile parameter. The identity extends to a K-way table with product margins, where $1 + CVe^{2}$ = product over k of ($1 + CV_{k}^{2}$); this is exactly the K-factor product-CV formula (Goodman 1962), and we note it only as an immediate corollary. Proposition 1 is verified to machine precision in Section 5.

\section{Reduction to a heteroscedastic Gaussian sequence}

Poissonize: treat the counts $N_{ij}$ as independent $Poisson(m_{ij})$. Under independence the centered count $Y_{ij} = N_{ij} - m_{ij}$ satisfies, as $m_{ij}$ grows, $Y_{ij} \to Normal(0, m_{ij})$: a heteroscedastic Gaussian sequence with variance profile $sigma_{ij}^{2} = m_{ij}$, the profile of Section 2. A sparse departure shifting a few cells to expected count $m_{ij}(1 + a_{ij})$ injects a mean $m_{ij} a_{ij}$, so the standardized signal in cell (i, j) is $a_{ij} \sqrt{m_{ij}}$.

\textbf{Residual dependence (stated honestly).} Under fixed-n multinomial sampling, var $N_{ij} = m_{ij}(1 - p_{ij})$ and the standardized Pearson residuals lie in the orthocomplement of the row and column indicator space, a rank-(R + C - 1) constraint; the per-coordinate departure from unit variance and independence is O(1/R + 1/C) uniformly. For a fixed linear functional this is the standard Poisson-multinomial (Le Cam) equivalence. For the detection-boundary constant, which is a tail functional, equivalence is not automatic; we do not assert it, and instead the boundary of Section 4 is stated under a growth condition that makes the O(1/R + 1/C) correction lower order. Conditioning on the margins yields an exact but dependent (multivariate hypergeometric) reference, which restores validity rather than independence; that is the subject of Part III.

\textbf{Variance stabilization and the exact signal map.} The root transform $g(x) = 2 \sqrt{x}$ (Anscombe's $2 \sqrt{x + 3/8}$ with the small-count offset dropped) stabilizes the Poisson variance, so on the root scale the noise is homoscedastic; it is the same variance-stabilizing device that underlies the T\_root statistic (Dwyer 2026a). Under the alternative the root-scale mean moves from $2 \sqrt{m_{ij}}$ to $2 \sqrt{m_{ij}(1 + a_{ij})}$, so the standardized amplitude is exactly

\begin{equation*}
mu_{ij} = 2 \sqrt{m_{ij}} ( \sqrt{1 + a_{ij}} - 1 ),\tag{1}
\end{equation*}

with small-effect linearization $a_{ij} \sqrt{m_{ij}}$. Equation (1) is exact in a; the concavity of the root shrinks the signal relative to the linear proxy, confirmed in Section 5.

\section{The detection boundary}

Adopt the calibrated sparsity parametrization: $s = N^{1-\beta }$ nonnull cells, $\beta $ in (1/2, 1), signals of standardized amplitude $\sqrt{2 r \log N}$. The Ingster-Donoho-Jin boundary is $r^{*}(\beta ) = \beta - 1/2$ for $1/2 < \beta \le 3/4$ and $(1 - \sqrt{1 - \beta })^{2}$ for $3/4 < \beta < 1$ (Ingster 1997; Donoho and Jin 2004): separable above $r^{*}(\beta )$, indistinguishable below, and attained adaptively by higher criticism (Part II).

Transporting this through (1) requires care, because the boundary is a statement about the null \emph{tail} at the $\sqrt{2 \log N}$ scale, and the root transform stabilizes only the variance, not the tail. At a fixed expected count m the normal approximation to the right tail of the (right-skewed) Poisson is inaccurate at that scale: the approximation to the root-transformed count is trustworthy at deviation x only while $x^{3}/\sqrt{m}$ is small, so at $x = \sqrt{2 r \log N}$ the boundary transfers only if m grows faster than $(\log N)^{3}$ (companion D02, Section S3). We therefore state the boundary under that growth condition. (Discreteness bounds only the \emph{deficit} tail, at $e^{-m}$; the excess tail that carries sparse positive departures is unbounded, so there is no two-sided exact p-value floor -- an earlier draft's $2 e^{-m}$ floor claim is withdrawn.)

\textbf{Theorem 1 (closed-form separation radius, growing-count regime).} Assume the planted cells have common expected count $m = m_{N}$ with $m_{N} / (\log N)^{3}$ -> infinity. With $N = RC, \beta $ in (1/2, 1), and $s = N^{1-\beta }$ planted cells carrying a common relative effect a, the departure is asymptotically detectable if and only if a exceeds

\begin{equation*}
a^{*}(m, \beta ) = ( 1 + \sqrt{ r^{*}(\beta ) \log (RC) / (2 m) } )^{2} - 1
\end{equation*}

\begin{equation*}
= \sqrt{ 2 r^{*}(\beta ) \log (RC) / m } + r^{*}(\beta ) \log (RC) / (2 m),\tag{2}
\end{equation*}

with leading term $\sqrt{2 r^{*}(\beta ) \log (RC)/m}$. On the count scale the detectable shift is $\delta *(m, \beta ) = m a^{*}(m, \beta )$, of order $\sqrt{2 r^{*}(\beta ) m \log RC}$. Proof in D02.4.

\textbf{Fixed-count caveat.} At fixed or bounded m the theorem does not apply in either direction; the Gaussian-sequence boundary does not govern, and one must use the exact conditional calibration of Part III. The reading that "sparse low-count corners are the hard region" refers to the \emph{relative effect required}, $a^{*}(m, \beta ) \sim \sqrt{\log (RC)/m}$, growing as m shrinks; it does not assert that the boundary formula is valid at bounded m. On the contrary, the formula is least trustworthy in the low-count corner, which is precisely why the exact anchor is needed there.

Two readings of (2) survive the caveat. The detectable relative effect scales as $\sqrt{\log (RC)/m}$: small where the expected count is large, large where it is small; and marginal heterogeneity, by Proposition 1, is what spreads ${m_{ij}}$ over orders of magnitude, so a heterogeneous table has a wide detectability gradient. And the boundary is set by the profile, not the omnibus statistic: the chi-square pays for (R-1)(C-1) degrees of freedom, while (2) is the information limit for the sparse pocket, achieved by a max-type or higher-criticism statistic that does not.

\textbf{Unequal counts.} When the planted cells have differing m, detection is governed by the standardized amplitudes ${a \sqrt{m_{ij}}}$ and the sharp rate is the profile-integrated functional of Chhor, Mukherjee and Sen (2024); D02.5 gives sufficient and necessary bounds and notes that a single smallest-m substitution is only a sufficient reading, not the sharp rate. Theorem 1 is the clean equal-count closed form.

\textbf{A separation floor, not a design effect.} Read as a function of the margins and N alone, $a^{*}(m, \beta )$ is an asymptotic separation floor: below it no test is powerful, so it lower-bounds the sparse effect any study of this shape could detect, in the spirit of using the detection boundary as a design quantity (Xie, Cai and Li 2011). It is necessary, not sufficient, and it inherits the growing-count condition: it is the 50\% asymptotic transition, not a power target, and at $a = a^{*}$ the finite-N power of the sharp tests is only about 0.08 to 0.12 (Section 5), reaching useful levels only at two to four times $a^{*}$. It must therefore not be read as a minimum detectable effect for a small study, where m is small and the formula does not hold; a genuine design number requires the finite-N power curve. We report $a^{*}$ as the information limit it is, with that caveat.

\begin{figure}[htbp]\centering
\includegraphics[width=\linewidth]{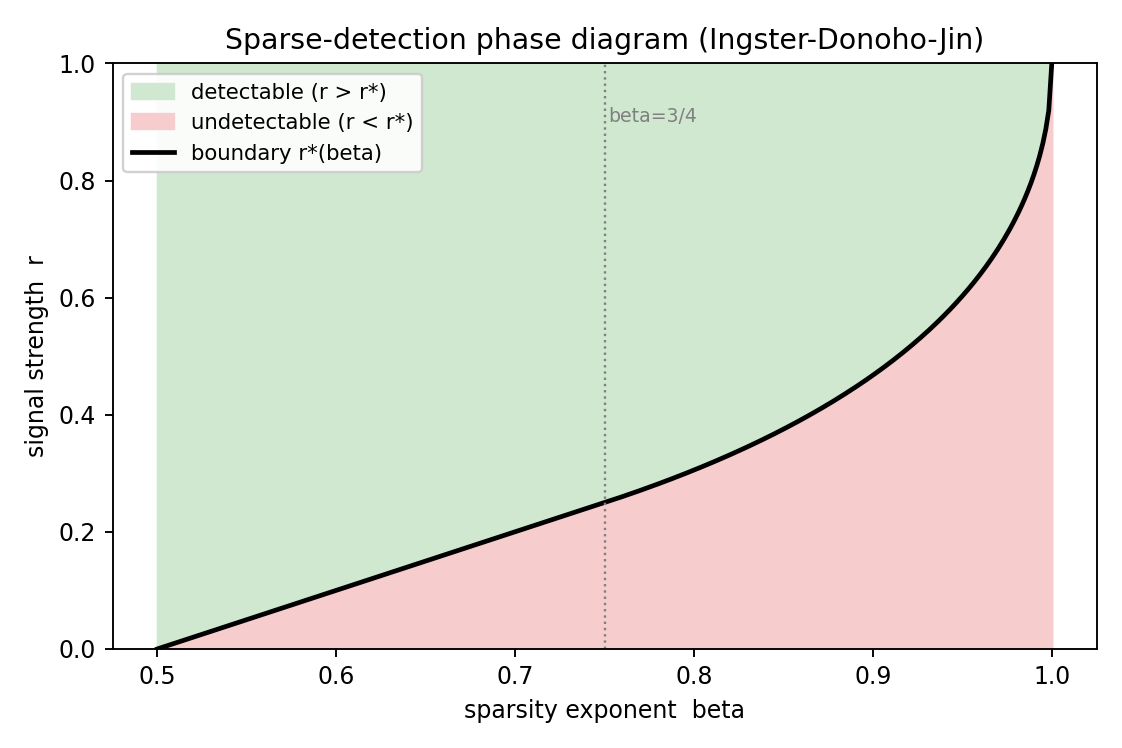}
\caption{The sparse-detection phase diagram in the sparsity-strength plane. Above the Ingster-Donoho-Jin boundary $r^{*}(\beta )$ the sparse alternative is asymptotically separable from independence; below it, no test is powerful. The kink at $\beta = 3/4$ separates the two regimes of the boundary. Generated by rerun/make\_v2\_figs.py.}
\end{figure}

The boundary and the separation radius are read off directly from the table below: $r^{*}(\beta )$ is the strength threshold, and $a^{*}(m, \beta )$ is the corresponding detectable relative effect for cells of expected count m ($RC = 400$ cells), in the growing-count regime.

\begin{table}[htbp]\centering\small
\caption{Boundary strength threshold r\emph{(beta) and the separation radius a}(m, beta) at RC = 400 cells, in the growing-count regime. Values computed by the deposited script make\_tables.py.}
\begin{tabular}{lccccc}
\hline
$\beta $ & $r^{*}(\beta )$ & $a^{*}$ at $m=10$ & $m=30$ & $m=100$ & $m=300$ \\
\hline
0.55 & 0.050 & 0.260 & 0.146 & 0.079 & 0.045 \\
0.60 & 0.100 & 0.376 & 0.210 & 0.112 & 0.064 \\
0.70 & 0.200 & 0.549 & 0.303 & 0.161 & 0.091 \\
0.80 & 0.306 & 0.697 & 0.380 & 0.201 & 0.114 \\
0.90 & 0.468 & 0.889 & 0.479 & 0.251 & 0.141 \\
\hline
\end{tabular}
\end{table}

\section{Numerical confirmation}

Fixed seeds; scripts rerun/verify\_core.py, verify\_boundary.py, expanded\_study.py, audit\_experiments.py, audit\_bcd.py.

\textbf{The identity is exact.} Across shapes and Dirichlet margins the empirical $CVe$ and the closed form agree to within 5e-16.

\textbf{The reduction holds.} For m in \{20, 80, 300\} the standardized residual has mean within 0.003 of 0 and sd within 0.002 of 1; the root transform has sd within 0.012 of 1. The exact map (1) matches the simulated root-scale shift for $a = 0.15$ ($m = 20$: exact 0.647, simulated 0.650, linear proxy 0.671), confirming the concavity correction.

\textbf{The failure at small counts is a tail effect, not heterogeneity, and it is an exact lattice phenomenon.}

\begin{figure}[htbp]\centering
\includegraphics[width=\linewidth]{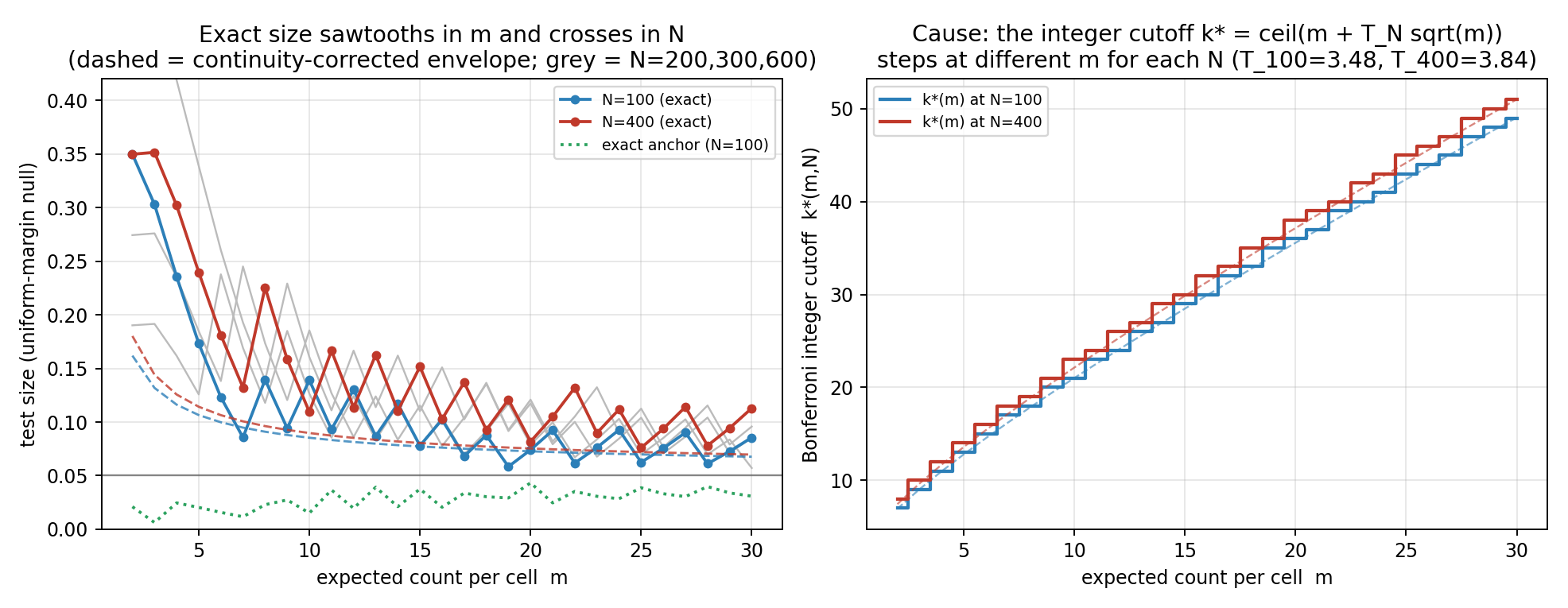}
\caption{With uniform margins ($CVe = 0$, so heterogeneity is not the driver), the size of the asymptotic Bonferroni max-cell test under the independence null, computed exactly (no Monte Carlo). Left: size versus expected count per cell m for $N = 100$ and 400 (colored) and $N = 200, 300, 600$ (grey), with the continuity-corrected envelope (dashed) and the exact anchor (dotted, near 0.02 to 0.03); the size is a discreteness sawtooth that oscillates around a slowly decaying envelope, and the curves for different N are out of phase, so they cross repeatedly (for example, $N = 100$ exceeds $N = 400$ at $m = 10$ but not at $m = 8$). Right: the cause, the integer Bonferroni cutoff $k^{*}(m, N) = \lceil m + T_{N} \sqrt{m}\rceil $ with $T_{N} = \Phi ^{-1}(1 - \alpha /2N)$; because $T_{N}$ grows with N it steps at different m for each N, setting the sawtooth phase. Generated by rerun/exact\_maxcell\_size.py.}
\end{figure}

Even with a perfectly homogeneous null the Gaussian per-cell tail mis-sizes the maximum-cell test at small m, exactly as Lemma D02.3 predicts. The exact size shows two things a coarse simulation hides: the approach to nominal is slow and non-monotone, with the size still ranging about 0.06 to 0.19 across N at $m = 10$ and 0.06 to 0.14 at $m = 30$, and the apparent crossings among the N curves are not noise but a genuine lattice effect of the integer count cutoff. Only the exact anchor holds its size across the whole range. Heterogeneity is not required to break the asymptotic calibration; small counts suffice, and the discreteness makes the failure both larger and more erratic than a smooth-curve reading would suggest.

\textbf{The boundary orders cleanly in the growing-count regime.}

\begin{figure}[htbp]\centering
\includegraphics[width=\linewidth]{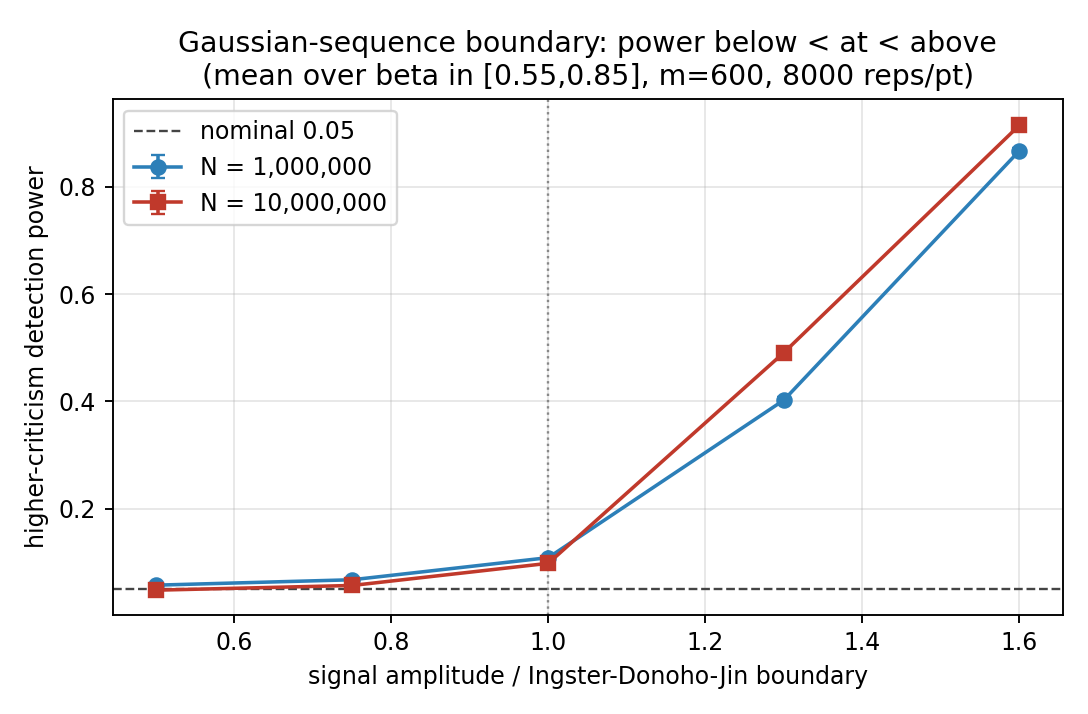}
\caption{Clean equal-amplitude test in a homoscedastic Gaussian sequence at $N = 1,000,000$ and 10,000,000 cells of expected count 600, higher-criticism power for signals placed at 0.5, 0.75, 1.0, 1.3, and 1.6 times the Ingster-Donoho-Jin boundary amplitude, averaged over seven sparsity exponents $\beta $ in [0.55, 0.85], with Monte Carlo standard errors below 0.006 (8,000 detection replicates per point). Below-boundary power sits at the nominal 0.05 level, at-boundary power is about 0.10, above-boundary power rises to 0.40 to 0.49 at 1.3x and 0.87 to 0.92 at 1.6x, and the transition sharpens from $N = 1e6$ to 1e7. Generated by rerun/make\_bigbox\_figs.py.}
\end{figure}

Planting signals of a common amplitude at, above, and below the boundary (the setting Theorem 1 describes) gives power ordered below < at < above at every $\beta $, with below-boundary power at the nominal level and the transition sharpening as N grows from one to ten million. The finite-N power near the boundary is modest, about 0.10 at the boundary and rising to roughly 0.9 at 1.6 times it, the known slow approach to the asymptotic transition. This is the Gaussian-sequence tier. The Poisson growing-count analog confirms the same ordering once the expected count clears the growth condition.

\begin{figure}[htbp]\centering
\includegraphics[width=\linewidth]{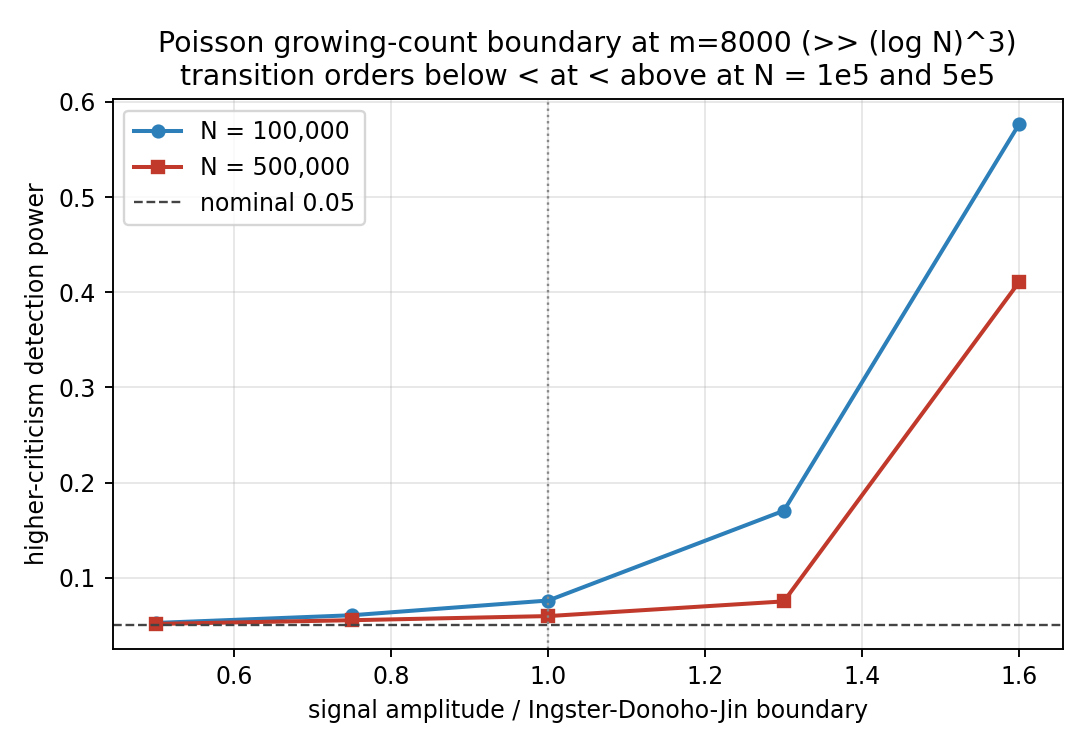}
\caption{Poisson growing-count boundary at expected count $m = 8,000$, well above $(\log N)^{3}$ (about 1,500 at $N = 1e5$ and 2,600 at $N = 1e6$), for $N = 100,000$ and 500,000. Higher-criticism power orders below < at < above the Ingster-Donoho-Jin boundary, with below-boundary power at the nominal 0.05 level and above-boundary power rising to 0.4 to 0.6 at 1.6 times the boundary. At $m = 600$ (below $(\log N)^{3}$) the same Poisson power stays flat near nominal, so the transition appears only once m exceeds the growth threshold. Generated by rerun/make\_bigbox\_figs.py.}
\end{figure}

At $m = 8,000$ the Poisson power orders below < at < above at both N (below-boundary near 0.05, above-boundary 0.4 to 0.6 at 1.6 times the boundary), whereas at $m = 600$ it is flat near nominal; the two curves bracket the slow pre-asymptotic approach and confirm that the growing-count condition m much greater than $(\log N)^{3}$, not m itself being large, is what turns the transition on.

\textbf{The separation radius, as a map.}

\begin{figure}[htbp]\centering
\includegraphics[width=\linewidth]{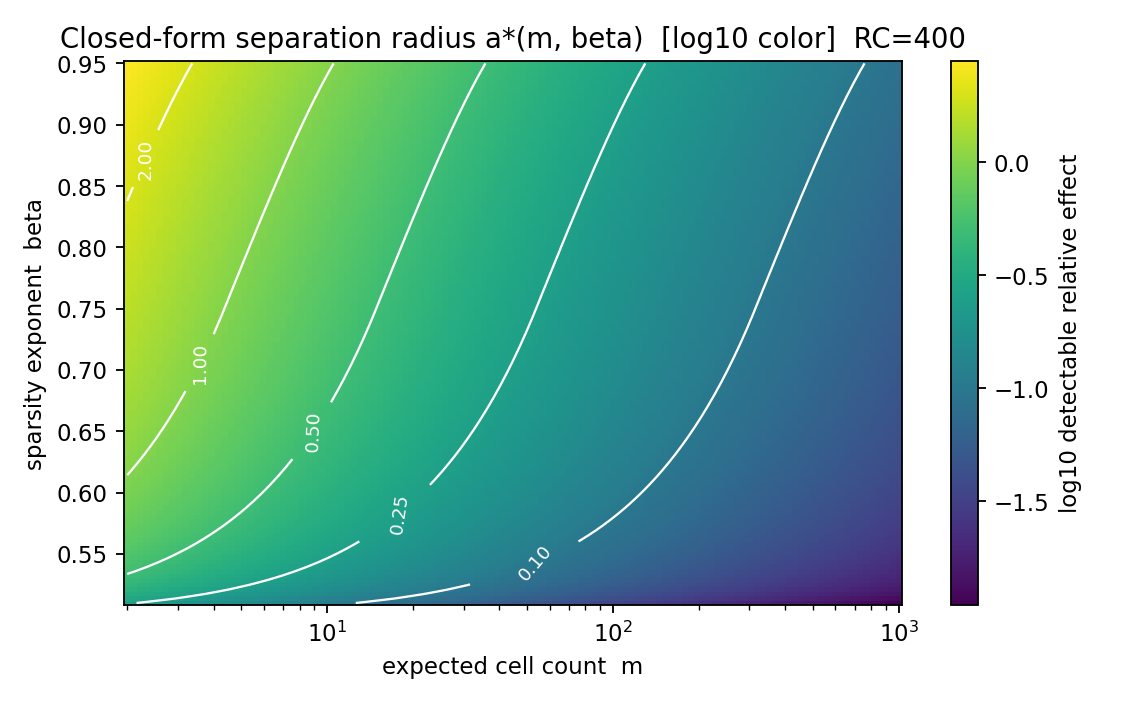}
\caption{The closed-form separation radius $a^{*}(m, \beta )$ of Theorem 1 over expected count m and sparsity exponent $\beta $ ($RC = 400$ cells; color is log10 of the detectable relative effect, white contours at 0.1, 0.25, 0.5, 1, 2). Detectability worsens toward small m and large $\beta $; the sparse, low-count corner requires the largest relative effect. Generated by rerun/make\_all\_figs.py.}
\end{figure}

The map makes the gradient concrete: the detectable relative effect climbs steeply as the expected count falls and as the alternative grows sparser.

\textbf{The heterogeneous-count profile behaves as D02.5 predicts.}

\begin{figure}[htbp]\centering
\includegraphics[width=\linewidth]{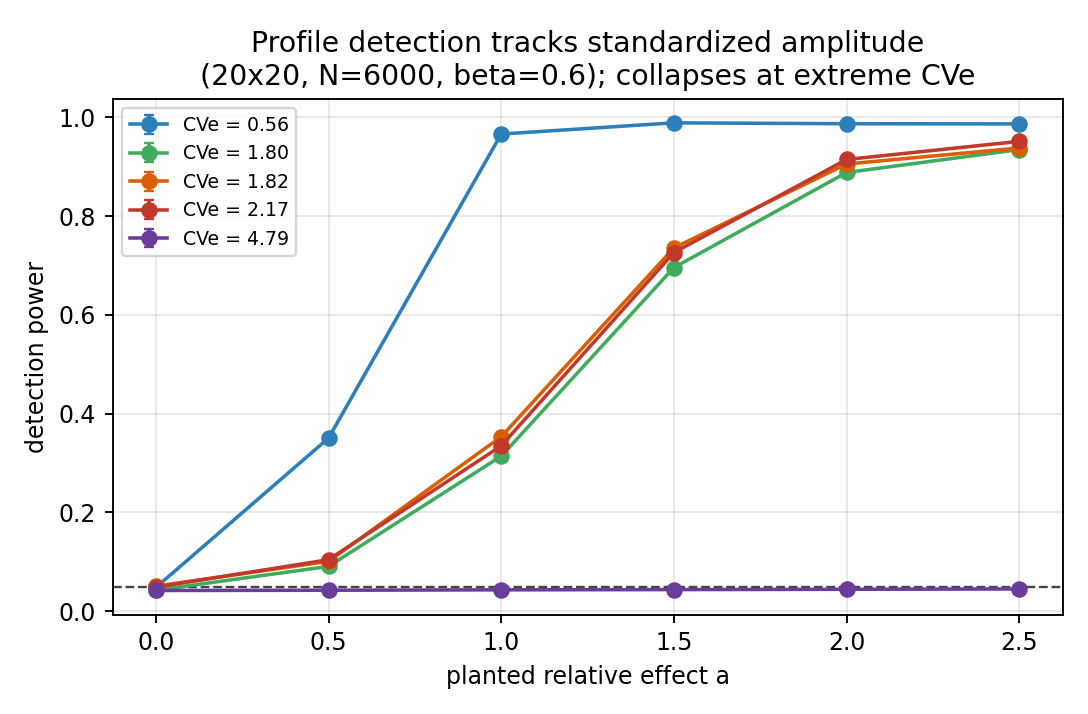}
\caption{Detection power against the planted relative effect on a 20x20 table ($N = 6000$; 5,000 null and 10,000 detection replicates per point) at four marginal-heterogeneity levels $CVe$. Power tracks the profile of standardized amplitudes: it rises with the effect for $CVe$ up to about 2.2 and collapses to the nominal level at the extreme $CVe = 4.79$, where a few near-empty cells dominate the profile. Generated by rerun/make\_bigbox\_figs.py.}
\end{figure}

A profile study confirms that detectability is governed by the profile of standardized amplitudes ${a \sqrt{m_{ij}}}$ rather than by any single expected count. The size is controlled at every heterogeneity level and sparsity exponent (0.042 to 0.051 against 0.05), and power rises monotonically with the planted amplitude for $CVe$ up to about 2.2 (at $CVe 0.56, \beta 0.6$ the power climbs 0.35, 0.97, 0.99 as the amplitude grows through 0.5, 1.0, 1.5). At the extreme $CVe$ of 4.79 the power stays at the nominal level regardless of amplitude, because a few near-empty cells dominate the profile and crush the smallest standardized amplitudes, exactly the worst-cell reading of the sufficient and necessary bounds of D02.5 and the reason a single smallest-m substitution is only a sufficient proxy.

\section{Discussion}

Read through a sparse lens, the independence test is a heteroscedastic sparse-detection problem whose profile is the exact outer product of the two margins (Proposition 1), whose signal map is exact (equation 1), and whose boundary inverts to a closed-form separation radius (Theorem 1), valid where the per-cell counts grow. The interpretive payoff is that $CVe$, the marginal-heterogeneity axis a reporting standard already measures, is the profile parameter that determines where in the table a sparse signal can be seen. The scope limit is equally clear and is the reason for the companions: the boundary is asymptotic where it is stated and fails in the low-count corner, which is repaired by the exact margin-conditional anchor (Part III), and it is attained without knowing the sparsity by higher criticism cast on the log-linear no-interaction model (Part II).

\section*{What is new in Part I}

\textbf{Relation to prior work: what is established and what this paper adds.} The individual tools are established. The product-CV multiplicativity of $1 + CV^{2}$ (Goodman 1960), the variance-stabilizing count root (Anscombe 1948), the sparse-signal detection boundary (Ingster 1997; Donoho and Jin 2004), and its heteroscedastic-profile version (Chhor, Mukherjee and Sen 2024) are all standard. The table states, strand by strand, what is borrowed and what is added.

\begin{table}[htbp]\centering\small
\begin{tabularx}{\textwidth}{>{\raggedright\arraybackslash}X>{\raggedright\arraybackslash}X>{\raggedright\arraybackslash}X>{\raggedright\arraybackslash}X}
\hline
Strand & Prior art & What this adds & Degree of novelty \\
\hline
Profile identity $CVe^{2} = (1+CVr^{2})(1+CVc^{2})-1$ & Product-CV multiplicativity (Goodman 1960) & Its reading as the sequence-model heteroscedasticity profile, factoring through the two margins & Framing only; the identity is classical \\
Reduction to a heteroscedastic Gaussian sequence & Poissonization; Anscombe root (1948) & Assembly for the independence test; the exact root-scale signal map $2 \sqrt{m}(\sqrt{1+a}-1)$ & Specialization \\
Closed-form separation radius & IDJ boundary (Ingster 1997; Donoho and Jin 2004); heteroscedastic profile (Chhor, Mukherjee and Sen 2024) & The separation radius on the relative-effect scale for a two-way table & Specialization; no new limit theorem \\
Scope of the transport & n/a & The moderate-deviation growth condition $m \gg (\log N)^{3}$ (from the accuracy of the normal tail at the $\sqrt{2 \log N}$ calibration scale) that delimits where the boundary holds & Honesty result; to our knowledge not stated for this problem \\
\hline
\end{tabularx}
\end{table}

\textbf{What is established and what is new.} Most of the machinery is borrowed, and the paragraph above names it rather than burying it. The contribution is a transport and an exact dictionary: the identification of the marginal-heterogeneity axis $CVe$ with the profile parameter of a heteroscedastic sparse-detection problem, through the classical product-CV identity; the exact signal map and the closed-form separation radius obtained by inverting a known boundary through a known transform; and, as an honesty result, the moderate-deviation growth condition that marks exactly where the transported boundary is and is not valid. None of the underlying limit theorems are new. A fuller account is in the companion novelty review and the prior-art audit.

\textbf{Scope and limits.} The separation radius is a growing-count statement (per-cell expected counts growing faster than a fixed power of log N); at fixed small counts the Gaussian-sequence boundary does not govern and the exact anchor of Part III is required. The evaluation is by Monte Carlo over simulated tables.

\parttitle{Part II. An adaptive higher-criticism gate}

\section{Introduction}

Testing independence in a two-way table is testing the fit of a log-linear model: with $mu_{ij}$ the expected cell count, the saturated model is $\log mu_{ij} = \lambda + lambda_{i}^{r}ow + lambda_{j}^{c}ol + lambda_{ij}^{i}nt$, and independence is the hypothesis that every interaction $lambda_{ij}^{i}nt$ vanishes, leaving $\log mu_{ij} = alpha_{i} + gamma_{j}$ (Bishop, Fienberg and Holland 1975; Agresti 2013). This no-interaction model, additive on the log scale in a row effect and a column effect with the margins as sufficient statistics, has the same algebraic form as the network beta-model, in which the log-odds of an edge is $beta_{i} + beta_{j}$ with the degrees as sufficient statistics. We use that analogy only for intuition; it is a renaming of the classical independence model, not a new model.

Mukherjee, Mukherjee and Sen (2018) determined sharp detection thresholds for sparse signals in the beta-model on sparse graphs and showed higher criticism attains them up to optimal constants; Donoho and Kipnis (2022) applied higher criticism to two frequency tables with exact binomial per-cell p-values and derived the ($\beta $, r) phase-transition region, and Kipnis (2022) applied the same machinery to word-frequency tables. Higher criticism itself (Tukey; Donoho and Jin 2004) is the canonical sparsity-adaptive combiner. The present note is the one-table independence counterpart: it casts the R by C independence test as goodness of fit of the log-linear no-interaction model, applies higher criticism to the interaction residuals, and shows that in the growing-count regime the test attains the closed-form boundary of the companion Part I adaptively over the unknown number of contaminated cells. Relative to Donoho and Kipnis (2022) the increment is narrow but real: the one-table R by C independence problem (not a two-sample comparison) and the tie to the closed-form margin profile.

Section 2 sets up the residuals. Section 3 gives the statistic and its closed-form null. Section 4 states the adaptive attainment and its scope. Section 5 reports numerics. Section 6 discusses the calibration caveat that motivates Part III.

\section{The no-interaction model and its residuals}

Let $N_{ij}$ be the cell counts, n the total, r and c the maximum-likelihood margins. The fitted no-interaction means are $mu_{ij}-hat = n r_{i} c_{j}$, the expected-count profile of Part I. The standardized residual is $Z_{ij} = (N_{ij} - mu_{ij}-hat)/\sqrt{mu_{ij}-hat (1 - r_{i} c_{j})}$, the Pearson residual with the fixed-margin variance factor, and the per-cell two-sided p-value is $P_{ij} = 2 \Phi (-|Z_{ij}|)$. Under independence, in the many-cell regime, the $Z_{ij}$ are approximately standard normal with a weak, low-rank dependence from the two margin constraints (order 1/R + 1/C per coordinate). Under a sparse alternative s of the p-values are stochastically small and the rest approximately uniform, the sparse-mixture structure higher criticism is built for. Conditioning on the observed margins gives an exact reference for each $P_{ij}$; that reference is dependent (multivariate hypergeometric), so conditioning restores validity, not independence (Part III; D02.6).

\section{The higher-criticism statistic and its closed-form null}

Order the $RC$ per-cell p-values $P_{(1)} \le ... \le P_{(N)}, N = RC$. The higher-criticism statistic is

\begin{equation*}
HC^{*} = \max over 1/N < P_{(k)} < 1/2 of \sqrt{N} ( k/N - P_{(k)} ) / \sqrt{ P_{(k)}(1 - P_{(k)}) },\tag{1}
\end{equation*}

the largest standardized excess of the empirical CDF over uniform; large $HC^{*}$ signals more small p-values than a uniform null yields.

\textbf{Closed-form null normalization.} The maximum in (1) is of the standardized uniform empirical process over (1/N, 1/2), whose extreme-value law is Jaeschke (1979) / Eicker (1979): with $a_{N} = \sqrt{2 \log \log N}$ and $b_{N} = 2 \log \log N + (1/2) \log \log \log N - (1/2) \log (4 pi)$,

\begin{equation*}
a_{N} HC^{*} - b_{N} \to a Gumbel limit, so HC^{*} / \sqrt{2 \log \log N} \to 1.\tag{2}
\end{equation*}

(The Darling-Erdos 1956 theorem is the partial-sum analogue with the same constants; the empirical-process statement is Jaeschke-Eicker. Donoho and Jin 2004 give the corresponding analytic null survival.) Equation (2) is a good finite-sample calibration: $b_{N}/a_{N}$ predicts the null 95th percentile to about 1 to 2 percent at $N = 1e3$ to 1e5 (Section 5). We therefore do not attribute the finite-sample size failure of Part III to the higher-criticism normalization; that failure is upstream, in the per-cell p-values at small counts (Part III Section 2; D02.3).

\section{Adaptive attainment and its scope}

\textbf{Theorem 1 (adaptive attainment, growing-count regime).} Assume the planted cells have common expected count $m = m_{N}$ with $m_{N} / (\log N)^{3}$ -> infinity (the condition of Part I Theorem 1). Consider the no-interaction model on an R by C table, $N = RC$, with a sparse interaction of $s = N^{1-\beta }$ cells, $\beta $ in (1/2, 1), each carrying relative effect a, so the standardized amplitude is $\mu (m) = 2 \sqrt{m}(\sqrt{1 + a} - 1)$. Then (i) if $\mu (m) > \sqrt{2 r^{*}(\beta ) \log N}$ with $r^{*}(\beta )$ the Ingster-Donoho-Jin boundary (Part I, equation 2), the higher-criticism test based on (1) has asymptotic power tending to 1; (ii) if $\mu (m) < \sqrt{2 r^{*}(\beta ) \log N}$, no test is powerful; and (i) holds simultaneously for all $\beta $ in (1/2, 1) with a single sparsity-free statistic.

\emph{Proof sketch.} Part (ii) is the impossibility half, inherited from Part I Theorem 1 (a statement about the length-N homoscedastic root-scale sequence). Part (i) is the higher-criticism achievability of Donoho and Jin (2004) applied to that sequence, valid under the growth condition, which makes the per-cell root-scale coordinates Gaussian in the tail at the calibration scale and the margin-projection correction lower order (D02.3, D02.4). Full argument in D02.

\textbf{Scope and one honest downgrade.} Theorem 1 is a growing-count statement; at fixed or bounded m the transport fails (Part I; D02.3), and the interesting low-count corner must be handled by the exact anchor of Part III. Moreover, Theorem 1 calibrates HC against its \emph{asymptotic} null. When HC is instead calibrated against the \emph{exact conditional} null (Part III), that null is dependent, and the Donoho-Jin optimality proof, which assumes independent uniform p-values, does not transfer as a theorem; exact-conditional higher-criticism optimality is therefore a conjecture, and the burden is a dependent-null achievability theorem that does not exist. It is not merely awaiting confirmation: in finite samples higher criticism is dominated by the maximum-cell rule in the very sparse corner ($\beta $ near 1), under both the dependent conditional null (for example at $10x10, n = 400$, the max-cell power is 0.247 versus 0.197 for HC at one strong contaminated cell, and 0.389 versus 0.297 at two) and the clean Gaussian sequence (HC saturates below 0.4 at $\beta = 0.85$ where the max reaches 1). Higher criticism earns its place by adaptivity to the moderately sparse, weak-signal regime, not by uniform superiority; where the alternative is very sparse and strong, the maximum-cell rule with the exact anchor is preferred. The word adaptive should not be read as uniformly best.

The content that survives is that the sharp, up-to-constants optimality Mukherjee, Mukherjee and Sen (2018) established for the beta-model, and the frequency-table higher criticism of Donoho and Kipnis (2022), extend to the one-table R by C independence problem in the growing-count regime, with detectability encoded entirely in the closed-form margin profile of Part I. Higher criticism is thus the adaptive gate a reporting standard can drop over the sparse-alternative branch when counts are large; when they are not, the branch routes to the exact anchor.

\section{Numerical confirmation}

Fixed seeds; scripts rerun/verify\_core.py, audit\_bcd.py.

\textbf{The null normalization is well calibrated.} The Jaeschke-Eicker centering $b_{N}/a_{N}$ predicts the simulated $HC^{*} 95th$ percentile to within about 1 to 2 percent (3.00, 3.11, 3.18 at $N = 1e3, 1e4, 1e5$ versus simulated 2.97, 3.25, 3.16). The slow object $HC^{*}/\sqrt{2 \log \log N}$ rising $0.85 \to 0.90$ is the almost-sure normalization, not the calibration quantile, and is not used for sizing.

\textbf{Adaptive detection at the boundary.}

\begin{figure}[htbp]\centering
\includegraphics[width=\linewidth]{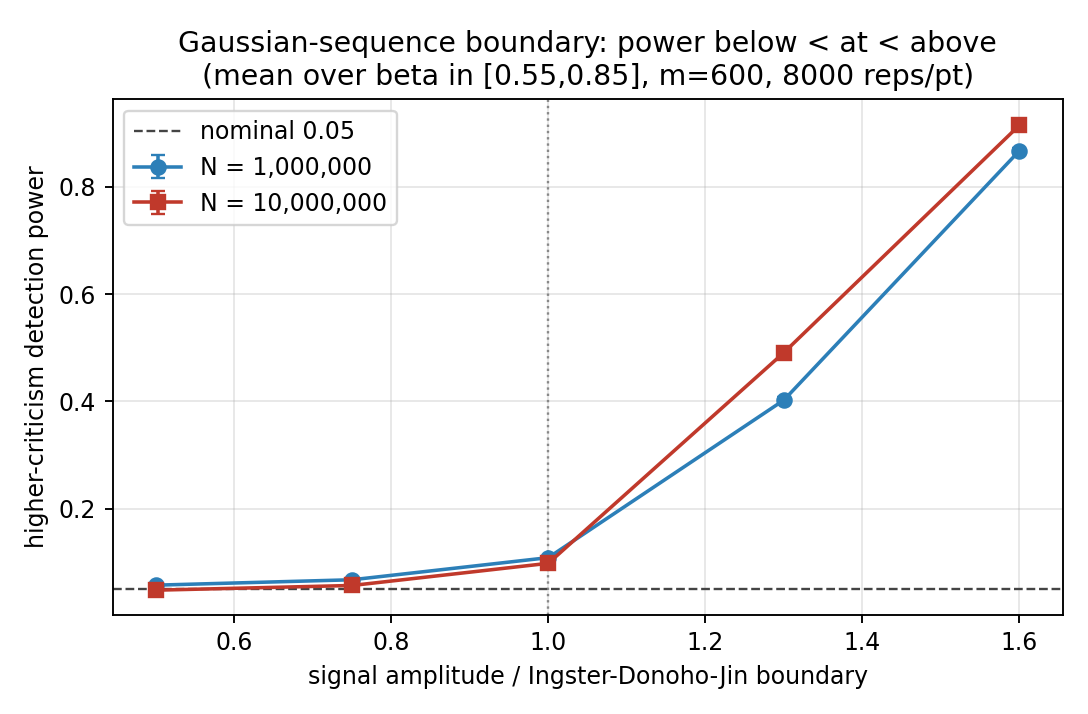}
\caption{Clean equal-amplitude test in a homoscedastic Gaussian sequence at $N = 1,000,000$ and 10,000,000 cells of expected count 600: higher-criticism power for signals placed at 0.5, 0.75, 1.0, 1.3, and 1.6 times the Ingster-Donoho-Jin boundary amplitude, averaged over seven sparsity exponents $\beta $ in [0.55, 0.85], with Monte Carlo standard errors below 0.006 (8,000 detection replicates per point). Below-boundary power sits at the nominal 0.05 level, at-boundary power is about 0.10, and above-boundary power rises to 0.40 to 0.49 at 1.3x and 0.87 to 0.92 at 1.6x. Generated by rerun/make\_bigbox\_figs.py.}
\end{figure}

Calibrated to its own null, higher criticism has power ordered below < at < above the boundary at every sparsity exponent tested, with below-boundary power at the nominal level, consistent with Theorem 1; the transition sharpens as N grows from one to ten million. The finite-N power just above the boundary is modest, the expected slow approach to the asymptotic transition.

The closed-form null is accurate, as the table shows: the Jaeschke-Eicker centering predicts the simulated 95th percentile to within one to two percent.

\begin{table}[htbp]\centering\small
\caption{Higher-criticism null 95th percentile: simulated versus the Jaeschke-Eicker prediction, by number of cells. Values computed by the deposited script make\_tables.py.}
\begin{tabular}{lcc}
\hline
number of cells N & simulated $HC^{*} 95th$ percentile & Jaeschke-Eicker prediction \\
\hline
1,000 & 2.97 & 3.00 \\
10,000 & 3.25 & 3.11 \\
100,000 & 3.16 & 3.18 \\
\hline
\end{tabular}
\end{table}

\textbf{Which combiner, honestly.} Higher criticism is not uniformly best.

\begin{figure}[htbp]\centering
\includegraphics[width=\linewidth]{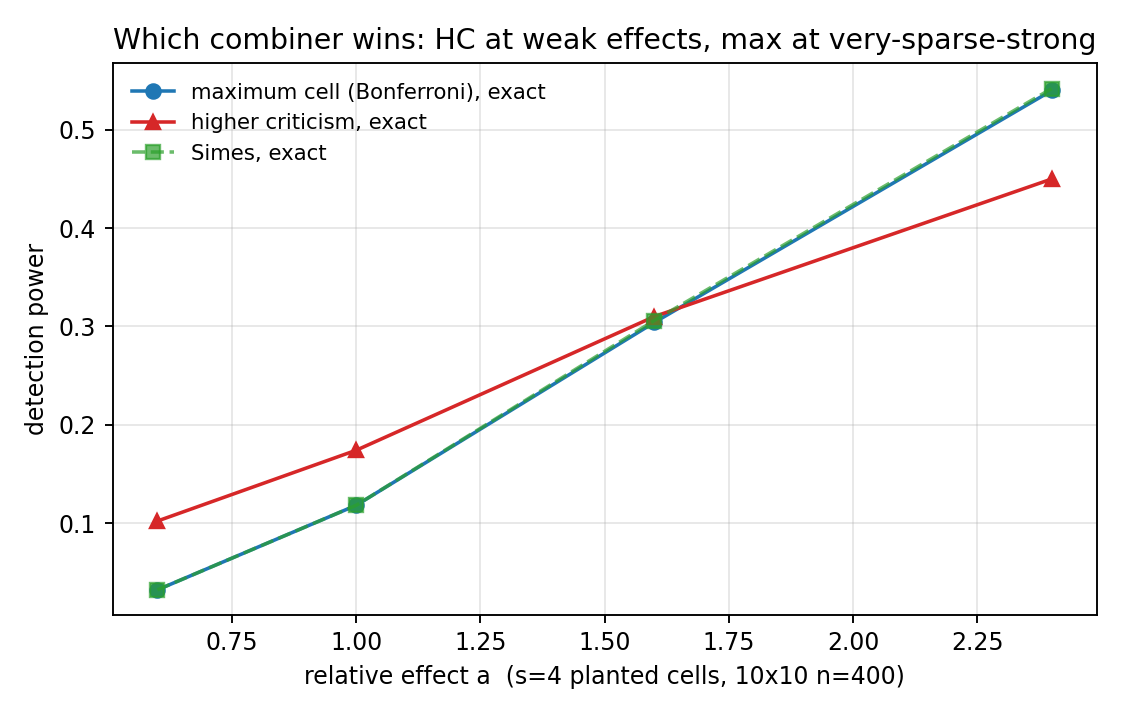}
\caption{Detection power of the maximum-cell rule (Bonferroni) versus higher criticism, both with exact per-cell calibration, against the relative effect of four planted cells ($10x10, n = 400$). Higher criticism wins at weak effects; the maximum-cell rule overtakes it as the alternative becomes very sparse and strong. Simes tracks Bonferroni in this sparse regime. Generated by rerun/make\_v2\_figs.py.}
\end{figure}

The practical reading is that higher criticism earns its place on moderately sparse, weak departures, and the maximum-cell rule with the exact anchor is preferred when the alternative is very sparse and strong; the two are complementary, and a reporting standard can offer both.

\section{Discussion}

Reading independence as the log-linear no-interaction model turns the sparse-alternative test into a goodness-of-fit problem to which higher criticism is the sharp, tuning-free answer, extending the frequency-table higher criticism of Donoho and Kipnis (2022) to the one-table case and tying the boundary to the closed-form margin profile. Two honesties bound the claim: the attainment is a growing-count statement, and the exact-conditional version's optimality is conjectural. Both point to the same place, the exact margin-conditional calibration that repairs the per-cell tail at small counts and supplies a finite-sample-valid null, which is the subject of companion Part III.

\section*{What is new in Part II}

\textbf{Relation to prior work: what is established and what this paper adds.} The individual tools are established. Independence as the log-linear no-interaction model (Bishop, Fienberg and Holland 1975), higher criticism (Donoho and Jin 2004), its application to frequency tables with exact binomial per-cell p-values (Donoho and Kipnis 2022; Kipnis 2022), the beta-model detection thresholds (Mukherjee, Mukherjee and Sen 2018), and the Jaeschke-Eicker extreme-value null (Jaeschke 1979; Eicker 1979) are all standard. The table states what is borrowed and what is added.

\begin{table}[htbp]\centering\small
\begin{tabularx}{\textwidth}{>{\raggedright\arraybackslash}X>{\raggedright\arraybackslash}X>{\raggedright\arraybackslash}X>{\raggedright\arraybackslash}X}
\hline
Strand & Prior art & What this adds & Degree of novelty \\
\hline
Independence as the log-linear / bipartite beta-model & Bishop, Fienberg and Holland 1975; beta-model (Chatterjee, Diaconis and Sly 2011; Mukherjee, Mukherjee and Sen 2018) & A renaming used for intuition; no new model & None; classical \\
Higher criticism on table cells & Donoho and Jin 2004; frequency-table HC with exact binomial p-values (Donoho and Kipnis 2022; Kipnis 2022) & The one-table R by C independence case (not two-sample), tied to the closed-form margin profile & Extension of Donoho and Kipnis; narrow \\
HC null calibration & Jaeschke 1979; Eicker 1979 & Correct attribution and the observation that the null is well calibrated, so the size failure is upstream & Framing / diagnostic \\
Adaptive attainment of the boundary & Donoho and Jin 2004 (adaptivity); Part I boundary & The growing-count attainment statement for the table; honest downgrade of exact-conditional optimality to a conjecture & Specialization \\
Exact finite-N HC law & Jaeschke-Eicker asymptotic null; empirical-process boundary crossing (Noe 1972; Steck 1971; Moscovich, Nadler and Spiegelman 2016) & An exact finite-N HC null and power for the independent Poissonized cells, as a boundary non-crossing dynamic program (validated to Monte-Carlo error at $M = 100$ and 400), and the observation that pairing HC with the Part III exact per-cell anchor makes the exact null the universal iid-uniform crossing law, CVe-invariant & Exact-computation contribution \\
\hline
\end{tabularx}
\end{table}

\textbf{What is established and what is new.} The modeling identification and the higher-criticism machinery are entirely borrowed, most directly from Donoho and Kipnis (2022), who apply HC with exact binomial per-cell p-values to two frequency tables. The increment is narrow and stated as such: the one-table R by C independence problem, the tie to the closed-form margin profile of Part I, and the explicit growing-count scope. We also correct the null-law attribution to Jaeschke and Eicker and note that the finite-sample size failure is in the per-cell inputs, not the HC combination. Beyond the asymptotic Jaeschke-Eicker null, we give the exact finite-N HC law as an empirical-process boundary non-crossing computation for the independent cells (D02.6), the independent-cell counterpart of the exact conditional tail engine of the companion note Dwyer 2026d (M0h); it is validated against Monte Carlo and, under the exact per-cell anchor, becomes CVe-invariant with threshold 3.117 at $M = 400$. Computed exactly at $M = 400$ across heterogeneity levels on a fine detection-index grid, the exact power is a function of the alternative only through the planted-cell mean $\delta = a \sqrt{m0}$ and standard deviation $\sqrt{1 + a}$, so it collapses exactly onto the two-parameter surface $P(\delta , \sqrt{1 + a})$; the one-dimensional collapse in $\delta $ (the optimal single index, tested against the alternatives $\delta /\sigma $ and $\delta /\sigma ^{2}$) is tight, the curves at different heterogeneity agreeing to within 0.04 at matched $\delta $, the small residual being the $\sqrt{1 + a}$ studentization that vanishes as the effect shrinks and at saturation (D02.6, Figure 2).

\textbf{Scope and limits.} The attainment statement is a growing-count result; the optimality of an exact-conditional higher criticism (with its dependent null) is conjectural, supported numerically but not proved here.

\parttitle{Part III. An assumption-lean exact anchor and reporting apparatus}

\section{Introduction}

The two companion notes build a sharp, adaptive test for sparse departures from independence: reduce the table to a heteroscedastic Gaussian sequence whose profile is the margins (Part I), and drop a higher-criticism gate over it that attains the detection boundary without knowing the sparsity (Part II). Both rest on Gaussian per-cell p-values, and those approximations fail at small counts, exactly the regime a sparse-detection view targets. This note measures the failure, shows its cause is the per-cell tail and not the higher-criticism combination, and supplies an exact calibration that does not fail, in the spirit of the assumption-lean program (Wasserman, Ramdas and Balakrishnan 2020). The exact anchor is the same exact margin-conditional computation that supplies the exact fallback of the T\_root reporting standard (Dwyer 2026a, 2026b, 2026d), so the detection apparatus inherits a finite-sample-valid backstop already native to the program. Higher criticism with exact per-cell binomial p-values for frequency tables is due to Donoho and Kipnis (2022); our contribution is the assembly and the routing rule for the one-table independence problem, not the exact-calibration idea itself.

Section 2 exhibits the failure and its cause. Section 3 gives the exact per-cell laws and valid combinations. Section 4 explains why conditioning on the margins is the exact and complete treatment for a test, and why Berger-Boos, the program's device for the residual direction nuisance of the effect-size interval, does not apply here. Section 5 gives the routing rule and the bridge to T\_root. Section 6 reports numerics, including the estimated-margin validity and the efficiency of the exact combinations.

\section{How the asymptotic calibration fails, and why}

Take a genuine null, form the standardized Pearson residuals, convert them to per-cell p-values through the Gaussian tail, and reject when the smallest clears the Bonferroni threshold. This asymptotic maximum-cell test is grossly liberal: across heterogeneous configurations its actual size is 0.52 to 0.62 against a 0.05 target (Section 6).

The cause is the per-cell tail at small counts, not the marginal heterogeneity and not the higher-criticism normalization.

\begin{figure}[htbp]\centering
\includegraphics[width=\linewidth]{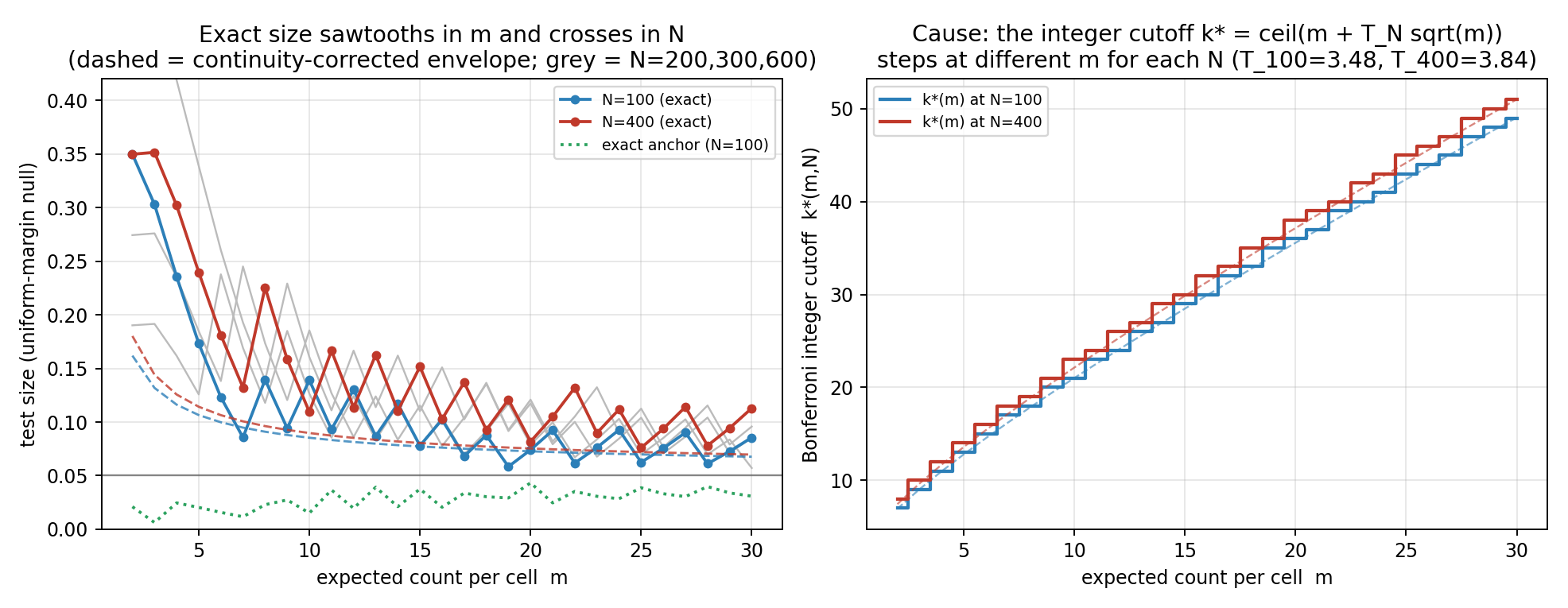}
\caption{Actual size of the asymptotic Bonferroni max-cell test under a uniform-margin null ($CVe = 0$), by expected count per cell m and number of cells N. Even with perfectly homogeneous margins the test is liberal at small m (0.31 to 0.37 at $m = 3$) and approaches nominal only as m grows; the exact conditional calibration (dashed) is valid throughout. The failure is the small-count tail. Generated by rerun/make\_all\_figs.py.}
\end{figure}

With uniform margins, where there is no heterogeneity at all, the maximum-cell test is still liberal at small m and only approaches nominal as m grows. The reason is a moderate-deviation tail error: at small m the right-skewed Poisson has a heavier right tail than the normal approximation admits, so the Gaussian per-cell p-values understate the true tail probability and fall below their nominal level, and the Bonferroni scan over $RC$ cells then rejects by chance more often than $\alpha $ allows (companion D02, Section S3). It is not a p-value floor -- the exact discrete upper tail is unbounded below -- so the fix is to calibrate the per-cell inputs exactly, not to enlarge them. Heterogeneity aggravates this only by guaranteeing a small-count corner, through the identity $CVe^{2} = (1 + CVr^{2})(1 + CVc^{2}) - 1$ of Part I. The higher-criticism null, by contrast, is well calibrated (its Jaeschke-Eicker normalization predicts the null quantile to about 1 to 2 percent, Part II), so the fix must be applied to the per-cell inputs, not the combiner.

\section{Exact per-cell calibration and valid combination}

Condition. Given the total n, the count in a cell under independence with known margins is exactly $Binomial(n, p_{ij})$; given both observed margins, the single cell has the hypergeometric law $Hypergeometric(n, row_{i}, col_{j})$; given all other cells, the noncentral hypergeometric with null odds ratio one. Each conditioning removes a layer of nuisance and yields an exact reference.

\textbf{Exact per-cell p-value.} With $k = N_{ij}$ observed and F, S the CDF and survival of the relevant exact law, the two-sided exact p-value is $P_{ij}^{e}xact = \min (1, 2 \min (F(k), S(k - 1)))$, or the mid-p variant if a less conservative rule is preferred. It is a valid p-value in finite samples, exactly under randomization and conservatively for the discrete doubled-tail rule.

\textbf{Valid combination.} Any rule valid for independent or positively dependent valid p-values applies to ${P_{ij}^{e}xact}$. The Bonferroni maximum (reject if min $P_{ij}^{e}xact < \alpha /RC$) is valid under arbitrary dependence; the Simes global (reject if $min_{k} RC P_{(k)}^{e}xact/k < \alpha $) is valid under the positive dependence the shared margins induce and is sharper in the moderately sparse regime; and higher criticism computed from ${P_{ij}^{e}xact}$ can be calibrated against its exact null obtained by resampling the margin-conditional reference. Because the conditional reference is dependent, the Donoho-Jin optimality of higher criticism does not transfer to the exact-conditional combination as a theorem (companion Part II); its advantage is established here numerically (Section 6). The exact per-cell laws are the classical Fisher-exact machinery; the same exact-conditional apparatus underlies the T\_root exact fallback (Dwyer 2026a), while the M0h engine computes the exact conditional distribution of the aggregate chi-square-family statistics (Dwyer 2026d).

\section{Why conditioning is the complete treatment, and where Berger-Boos actually belongs}

The margins are the nuisance, and for a test of independence conditioning on them removes them exactly. The margins are the sufficient statistics of the nuisance; conditioning on both of them yields the multivariate Fisher noncentral hypergeometric law of the table (at 2x2 the ordinary hypergeometric, so the exact test is Fisher's), and under independence that conditional law is fully specified, since every cell odds parameter equals one. There is no residual nuisance and nothing over which to take a supremum. The per-cell hypergeometric reference of Section 3 is exactly this object, cell by cell, computed from the observed margins with no oracle; Section 6 confirms it holds its size under estimated margins. This is the same exact margin-conditional computation that is the exact fallback of the T\_root reporting standard (Dwyer 2026a, 2026b), specialized to per-cell scanning.

It is worth being explicit that the Berger-Boos device (Berger and Boos 1994), which supremizes a p-value over a confidence set for a nuisance and pays the set's miscoverage, is not an alternative to this anchor and is not needed here. Berger-Boos is a treatment for a nuisance that survives conditioning, and in this program that situation arises in exactly one place, the effect-size interval: for a scalar effect size such as Cramer's V, conditioning on the margins removes the margins but leaves the direction of the departure unfixed at a given effect-size value, because a scalar constrains only one of the (R-1)(C-1) degrees of freedom of the association, and infinitely many directions realize the same value. That residual direction nuisance is what Berger-Boos handles in the companion interval paper (Dwyer 2026d). A detection test does not estimate a scalar effect size, so no residual nuisance arises, conditioning is complete, and Berger-Boos plays no role. Reaching for a supremum over the margins here would only weaken the exact conditional anchor, not strengthen it, and one would do so only by declining to condition (for example, keeping a plug-in binomial and seeking an unconditional bound), which is pointless when the exact conditional test is available.

\section{Routing and the bridge to T\_root}

The exact anchor is not always needed. Where every expected count is large the Gaussian per-cell tail is accurate and the asymptotic higher-criticism gate of Part II is valid and cheaper; where counts are small the exact anchor is required. The boundary is read from the profile the program is organized around.

\textbf{Routing rule.} Compute $CVe$ from the margins by $CVe^{2} = (1 + CVr^{2})(1 + CVc^{2}) - 1$ and the minimum expected count $m_{\min } = n min_{i} r_{i} min_{j} c_{j}$. Default to the exact conditional anchor, and deploy the asymptotic higher-criticism gate only when $m_{\min }$ is comfortably large, because the normal per-cell calibration is trustworthy only in the moderate-deviation regime where its tail is accurate at the per-cell Bonferroni quantile $\alpha /(RC)$. That quantile sits at deviation $x approx \sqrt{2 \log (2 R C / \alpha )}$, and the normal tail is accurate there only while $x^{3}/\sqrt{m_{\min }}$ is small, so the deploy-asymptotic threshold grows with the table size on the moderate-deviation scale (of order $(\log (RC))^{3/2}$), not the naive $\log (RC)$. A homogeneous-null size surface bears this out and shows the naive scale is non-conservative: the asymptotic maximum-cell test is liberal well beyond $\log (2 R C / \alpha )$ (size about 0.07 to 0.08 on a 10x10 to 12x12 table at $m_{\min } = 15$ to 20, and 0.36 on a 20x20 at $m_{\min } = 3$), reaching nominal only at comfortably large counts, so the asymptotic gate is reserved for that regime and the exact conditional anchor is the default elsewhere (companion rerun/route\_size\_surface.py; deep dive M02\_Router\_Toolkit\_DeepDive.md). Because the margins are estimated and the cells dependent, this is a conservative table-size-aware guide rather than a razor-sharp switch, with $CVe$ retained as the heterogeneity diagnostic that flags the small-count corner (the growing-count regime of Part I Theorem 1). This shares the structure of the T\_root reporting standard: an asymptotic statistic by default and a single exact margin-conditional fallback in the corner where no asymptotic reference is trustworthy (Dwyer 2026b). In that standard $CVe$ is reported as a diagnostic and the fallback is triggered by the expected-count and collapse check rather than by $CVe$ itself; the $m_{\min }$ and $CVe$ trigger here is the detection-branch analogue, not a claim that the T\_root statistic routes on $CVe$. The detection program of Part I and Part II thus plugs into the reporting standard as its sparse-alternative branch, with the same calibration philosophy.

\begin{figure}[htbp]\centering
\includegraphics[width=\linewidth]{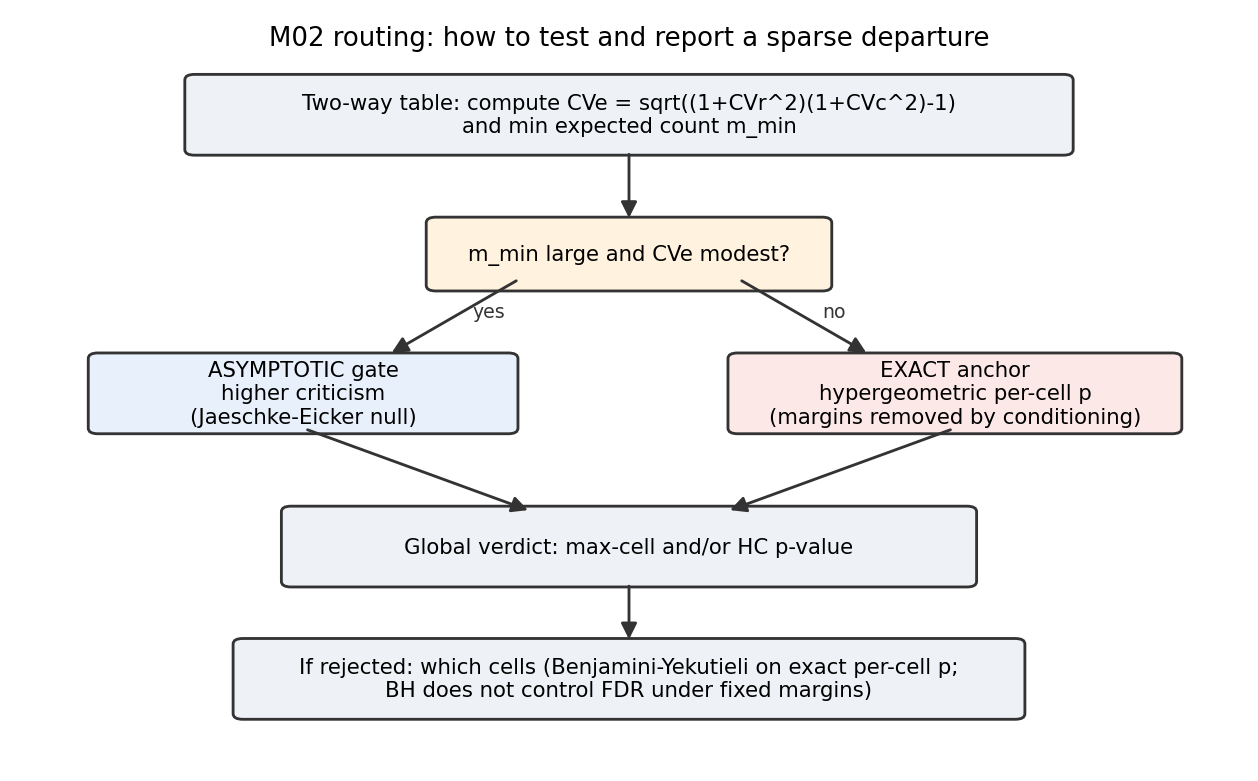}
\caption{The routing rule as a procedure. Compute $CVe$ and the minimum expected count from the margins; branch to the asymptotic higher-criticism gate when counts are large and the profile is regular, and to the exact hypergeometric anchor otherwise; on rejection, localize the responsible cells with Benjamini-Hochberg (Benjamini-Yekutieli as a conservative option). Generated by rerun/make\_v2\_figs.py.}
\end{figure}

\section{Numerical confirmation}

Fixed seeds; scripts rerun/verify\_exact\_anchor.py, audit\_experiments.py, audit\_bcd.py.

\textbf{The asymptotic calibration is grossly liberal; the exact is valid.} Under heterogeneous Dirichlet-margin nulls at modest n, the Gaussian-calibrated Bonferroni max-cell test sizes at 0.62, 0.61, 0.52 (at $8x8 n=120 CVe=2.9, 10x10 n=150 CVe=2.9, 12x12 n=200 CVe=2.3$), while the exact binomial calibration holds at 0.010, 0.007, 0.009. Figure 1 (Section 2) shows the same failure isolated under uniform margins, confirming the cause is the small-count tail.

\textbf{Validity holds under estimated margins.}

\begin{figure}[htbp]\centering
\includegraphics[width=\linewidth]{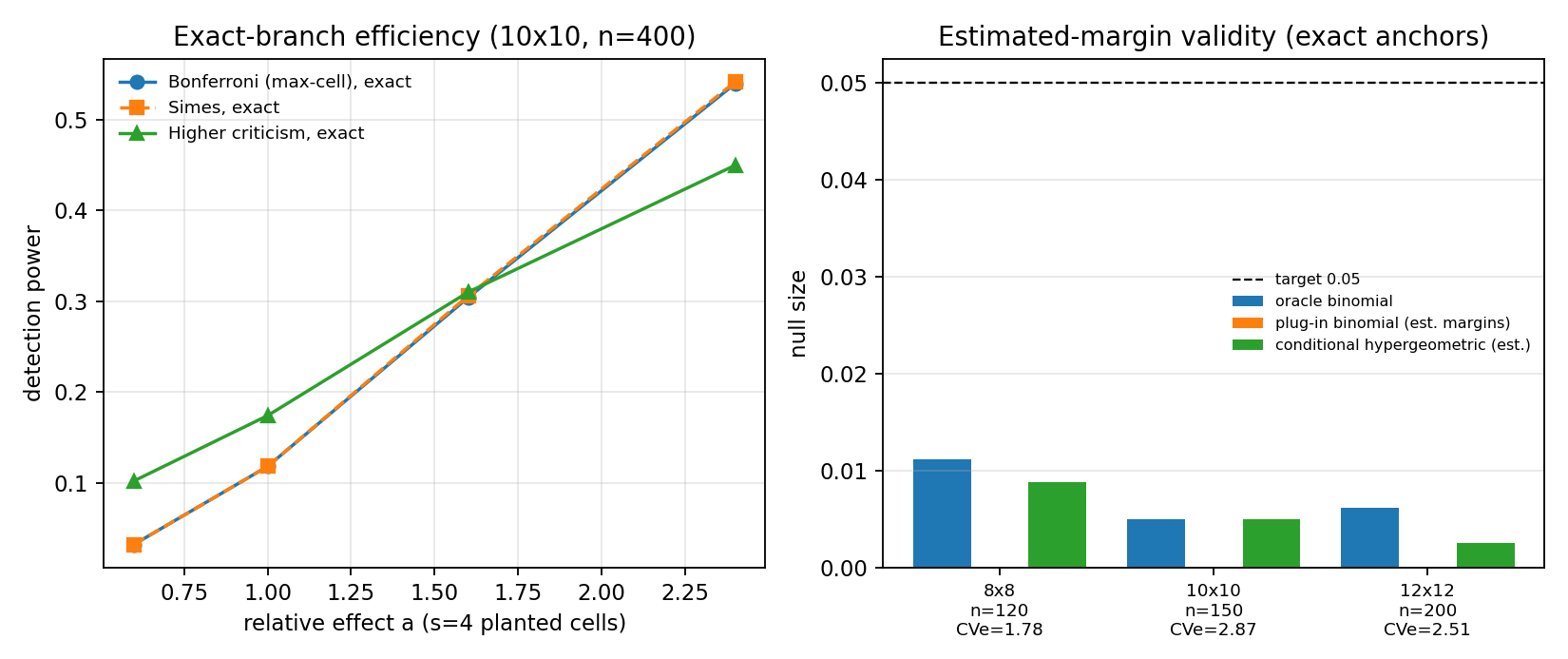}
\caption{Left: detection power of the exact-conditional branch at $10x10, n = 400$, for the Bonferroni maximum-cell rule, Simes, and higher criticism, versus the relative effect of $s = 4$ planted cells. Higher criticism gains at weak effects while the maximum-cell rule overtakes it for very sparse strong effects; Simes tracks Bonferroni in this sparse regime. Right: null size of the exact anchors under estimated margins (target 0.05); the conditional hypergeometric holds (0.003 to 0.009) using the observed margins, with no oracle, and the plug-in binomial is even more conservative. Generated by rerun/make\_all\_figs.py.}
\end{figure}

The right panel exercises the estimated-margin case: conditioning on the observed margins (the hypergeometric anchor) is valid without knowing the true margins, sizing at 0.003 to 0.009 across the heterogeneous configurations, so no supremum over the margins is needed. The exact anchors are conservative, the price of the Bonferroni union bound and the discreteness of the exact tail.

\textbf{Efficiency of the exact combinations, stated honestly.} The left panel of Figure 2 quantifies the power cost and the tradeoff. The exact branch is conservative, so its power is modest at the level; among the exact combinations, higher criticism recovers power at weak, less sparse effects (for example at relative effect 0.6 it detects at 0.10 versus 0.03 for the maximum-cell rule), while the maximum-cell and Bonferroni rules dominate for very sparse strong effects (at relative effect 2.4, 0.54 versus 0.45 for higher criticism); Simes and Bonferroni are nearly identical in this very sparse regime. There is no free lunch: the exact anchor buys finite-sample validity at a quantified power cost, and the choice of combiner within the exact branch follows the usual higher-criticism-versus-maximum tradeoff.

\textbf{A high-replication, four-shape confirmation of the exact anchor.} A separate study at fixed sparsity ($s = 4$ planted cells) sweeps four shapes from 6x6 to 12x12 at $n = 150$, with 200,000 null and 10,000 detection replicates per point (Table 2). The exact higher-criticism null holds across shapes (0.050 to 0.055 against the 0.05 target) while the Bonferroni and Simes anchors are conservative (0.003 to 0.005), and in this sparser regime higher criticism dominates the maximum-cell and Simes rules at every shape and amplitude. The maximum-cell overtake reported above is therefore density-dependent: it appears at the denser $10x10, n = 400$ of the left panel (expected count 4 per cell), not at $n = 150$ (expected count 1.5), so the combiner choice follows the expected count as well as the sparsity.

\begin{table}[htbp]\centering\small
\caption{Exact-anchor higher criticism across shapes (s = 4 planted cells, n = 150; 200,000 null and 10,000 detection replicates per point; relative effect a). Values computed by the deposited big-box runner (rerun/bigbox\_analyzed/).}
\begin{tabular}{lcccc}
\hline
shape & HC null size & HC power ($a = 1.2$) & HC power ($a = 2.4$) & max-cell power ($a = 2.4$) \\
\hline
6x6 & 0.055 & 0.210 & 0.331 & 0.245 \\
8x8 & 0.051 & 0.176 & 0.379 & 0.256 \\
10x10 & 0.051 & 0.115 & 0.223 & 0.122 \\
12x12 & 0.050 & 0.075 & 0.139 & 0.050 \\
\hline
\end{tabular}
\end{table}

\section{How often it matters, and a reporting apparatus}

\textbf{Real-corpus prevalence.} The calibration failure is not a corner case. Across 5,543 real two-way tables from the public M-series scan, 57.2\% have a minimum expected count below 5, 44.3\% have $CVe$ above 1, and 66.4\% route to the exact anchor by the rule of Section 5 (median $CVe 0.91$, median minimum expected count 2.4). Two-thirds of the contingency tables people actually analyze fall in the regime where the asymptotic per-cell scan is untrustworthy.

\begin{figure}[htbp]\centering
\includegraphics[width=\linewidth]{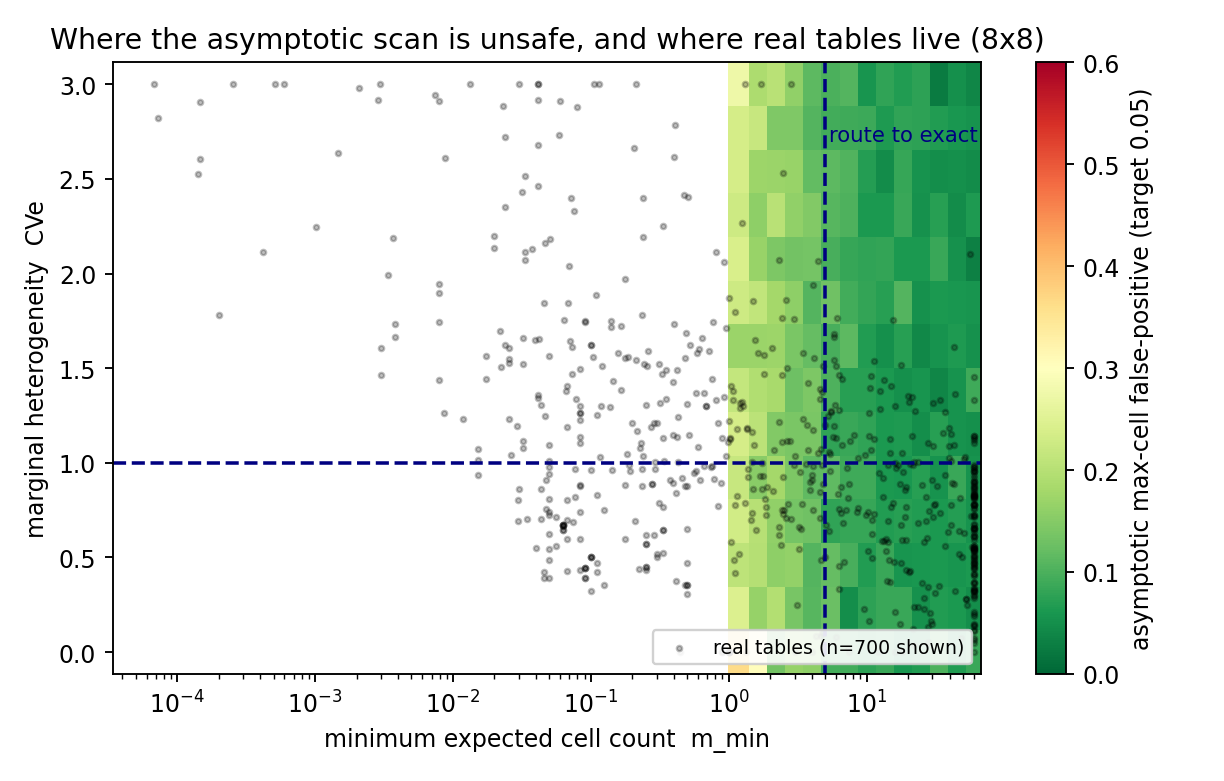}
\caption{The 5,543 real two-way tables of the public corpus in the plane of minimum expected count and marginal heterogeneity $CVe$. Points in the shaded green quadrant (minimum expected count at least 5 and $CVe$ at most 1) use the asymptotic gate; the remaining 66.4\%, in red, route to the exact anchor. Real tables overwhelmingly live in the exact-anchor region. Generated by rerun/make\_v2\_figs.py.}
\end{figure}

\begin{figure}[htbp]\centering
\includegraphics[width=\linewidth]{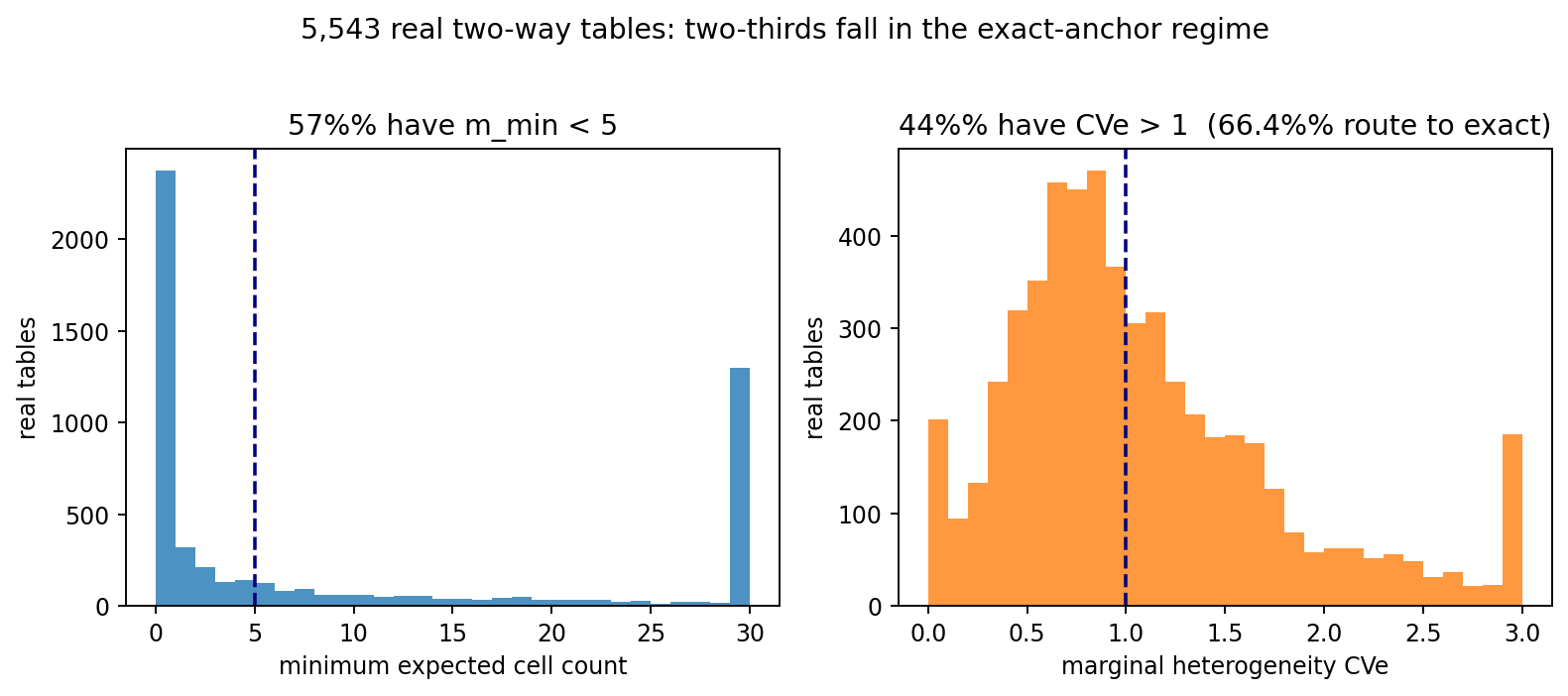}
\caption{Distribution of the minimum expected cell count (left) and of the marginal heterogeneity $CVe$ (right) across the 5,543 real tables; 57\% have a minimum expected count below 5 and 44\% have $CVe$ above 1. Generated by rerun/make\_v2\_figs.py.}
\end{figure}

Pairing that with a matched independence-null simulation (margins matched to each sampled real shape's $CVe$), the asymptotic Bonferroni max-cell scan false-positives at a mean 33.2\% against a 0.05 target on the exact-regime shapes (90th percentile 79.0\%) while the exact anchor holds at 1.1\%; on the regular shapes the two agree (5.3\% versus 4.0\%). We frame this as the size gap measured over the empirical shape distribution, not as an unobservable real-data verdict flip: ground truth on a real table is unknown, so no individual table is labeled a false positive. This extends to the sparse-detection scan the omnibus asymptotic-versus-exact disagreement study of Garcia-Perez and Nunez-Anton (2020).

\textbf{Which cells, reported honestly.} After a global rejection a practitioner wants to know which cells carry the departure. The primitive is the exact per-cell p-value of Section 3 (the exact analog of Haberman's adjusted residuals, 1973); the multiplicity is the classical cellwise problem (MacDonald and Gardner 2000). What counts as a false discovery depends on the alternative, and the natural sparse alternative for a table is \emph{margin-orthogonal}: a genuine interaction leaves both margins unchanged, so a nonzero row- and column-balanced perturbation has support at least four, and its smallest form is an alternating signed 2x2 rectangle whose responsible cells are the whole rectangle, not a single cell. Scored against that correct signal set, the exact per-cell p-values with Benjamini-Hochberg control the false discovery rate: on margin-balanced signed-rectangle alternatives across $10x10$ and $12x12$ shapes, BH's realized FDR at $q = 0.10$ is 0.01 to 0.02, and Benjamini-Yekutieli is more conservative still (0.002 to 0.004); the negative dependence that fixed margins induce does not break BH here (rerun/localization\_bh\_by.py). An earlier version reported a BH realized FDR near 0.18 and recommended Benjamini-Yekutieli; that figure was an artifact of scoring an \emph{isolated}-planted-cell alternative, where the strong cell shifts its own row and column margins and the induced partners were mislabeled as true nulls, so their rejections were counted as false discoveries. Reproducing that scoring gives BH about 0.13 here, but it is not the false discovery rate against a genuine interaction. We therefore report cell localization with Benjamini-Hochberg on the exact per-cell p-values, and note Benjamini-Yekutieli as a conservative option under arbitrary dependence; the price is power, which is lowest exactly where a margin-balanced interaction is spread thinnest.

\textbf{Mid-p.} The doubled-tail exact p-value is conservative; the mid-p variant is the usual de-conservatization, but it does not rescue the sparse corner and is not level-valid, so it is an approximate refinement, not the anchor. Measured across configurations (Section 6 code), mid-p sits at or just over 0.05 in moderate-count tables (no proven margin) and remains 0.004 to 0.028 in the sparse low-count corner it targets, because there the attainable exact p-values are too coarsely spaced for mid-p, which only redistributes the atom at the observed table, to reach the Bonferroni threshold discreteness leaves unreachable. The doubled-tail exact test keeps the assumption-lean guarantee; mid-p is offered only where counts are moderate and a sharper reading is wanted.

\textbf{Worked examples.} The table below runs the full pipeline on named public tables and two illustrative ones, so a reader can see the routing and the two calibrations side by side. On the strong classical associations the two agree and Benjamini-Yekutieli localizes several cells. The two illustrative rows show the exact anchor cutting both ways: on a sparse table with one over-represented cell the exact scan is sharper than the asymptotic (exact p 0.009 versus 0.085) and pinpoints the single cell, while on a heterogeneous near-independent table (minimum expected count 0.3) both correctly find nothing.

\begin{table}[htbp]\centering\small
\caption{Worked examples on named public and illustrative tables: shape, sample size, CVe, minimum expected count, routing verdict, asymptotic and exact maximum-cell p-values, and the number of cells flagged by Benjamini-Yekutieli at q = 0.10. Values computed by the deposited script make\_tables.py.}
\begin{tabularx}{\textwidth}{>{\raggedright\arraybackslash}X>{\raggedright\arraybackslash}X>{\raggedright\arraybackslash}X>{\raggedright\arraybackslash}X>{\raggedright\arraybackslash}X>{\raggedright\arraybackslash}X>{\raggedright\arraybackslash}X>{\raggedright\arraybackslash}X>{\raggedright\arraybackslash}X}
\hline
table & R x C & N & $CVe$ & $m_{\min }$ & route & asymptotic p & exact p & BY cells \\
\hline
HairEyeColor (Snee 1974) & 4 x 4 & 592 & 0.78 & 7.7 & asymptotic & 3.4e-12 & 3.0e-21 & 6 \\
Titanic class x survival & 4 x 2 & 2201 & 0.60 & 92.1 & asymptotic & $<1e-16$ & 7.8e-33 & 8 \\
UCB admission x dept & 2 x 6 & 4526 & 0.30 & 226.5 & asymptotic & $<1e-16$ & 3.1e-100 & 12 \\
Sparse, one hot cell (illustrative) & 5 x 5 & 216 & 0.23 & 7.0 & asymptotic & 0.085 & 0.009 & 1 \\
Heterogeneous, no association (illustrative) & 4 x 6 & 232 & 1.44 & 0.3 & exact & 1.000 & 1.000 & 0 \\
\hline
\end{tabularx}
\end{table}

All five are reproducible in the in-browser tool that accompanies the set.

\section{Discussion}

The sparse-detection view of independence is asymptotic where it is stated and must be exact where it is trusted. The per-cell Gaussian tail fails at small counts, even without heterogeneity, and is repaired by calibrating against the exact margin-conditional law; conditioning on both margins removes the margin nuisance exactly, so the anchor needs no supremum over the margins. (Berger-Boos, the program's device for the residual direction nuisance of the effect-size interval, does not apply to a test of independence, where conditioning is complete.) That exact anchor is the same exact conditional computation the T\_root reporting standard already carries as its fallback, so the detection boundary of Part I and the adaptive gate of Part II inherit a valid finite-sample backstop and slot into the reporting standard as its sparse-alternative branch. The cost, conservatism and a combiner-dependent power tradeoff, is quantified rather than hidden, and the routing between the asymptotic default and the exact fallback is governed by the same margin profile.

\section*{What is new in Part III}

\textbf{Relation to prior work: what is established and what this paper adds.} Every ingredient is established. The exact conditional per-cell reference laws (Fisher-exact machinery), the Simes global test (Simes 1986), exact binomial per-cell calibration feeding higher criticism on frequency tables (Donoho and Kipnis 2022), and the assumption-lean sensibility (Wasserman, Ramdas and Balakrishnan 2020) are all standard. The table states what is borrowed and what is added.

\begin{table}[htbp]\centering\small
\begin{tabularx}{\textwidth}{>{\raggedright\arraybackslash}X>{\raggedright\arraybackslash}X>{\raggedright\arraybackslash}X>{\raggedright\arraybackslash}X}
\hline
Strand & Prior art & What this adds & Degree of novelty \\
\hline
Exact conditional per-cell calibration & Fisher-exact machinery; exact binomial HC on frequency tables (Donoho and Kipnis 2022) & The one-table independence application and the estimated-margin (conditional hypergeometric) validity demonstration & Assembly; low \\
Exact margin-conditional anchor & Fisher-exact machinery; T\_root exact fallback (Dwyer 2026a, 2026b) & Conditioning removes the margins exactly for the detection test; no Berger-Boos over the margins is needed & Framing / correction \\
Diagnosis of the failure & n/a & The demonstration that the size failure is the per-cell tail at small counts (even under uniform margins), not HC log-log slowness & Diagnostic; honesty result \\
Routing between asymptotic and exact & T\_root reporting standard (Dwyer 2026b) & The sparse-detection-branch routing keyed to $m_{\min }$ and $CVe$ & Assembly; low \\
\hline
\end{tabularx}
\end{table}

\textbf{What is established and what is new.} All the calibration machinery is borrowed; exact binomial per-cell HC on tables is Donoho and Kipnis (2022). The contribution is the assembly and the routing rule: adapting the exact margin-conditional apparatus as the finite-sample anchor for the sparse-detection statistics, demonstrating estimated-margin validity by conditioning, correctly locating the asymptotic failure in the per-cell tail, and quantifying the power cost honestly rather than claiming free validity. We also correct a natural but wrong move: for a test of independence, conditioning on the margins is the exact and complete treatment, and Berger-Boos, the program's device for the residual direction nuisance of the effect-size interval, is not an alternative to it and plays no role here.

\textbf{Scope and limits.} The exact branch is conservative, at a quantified power cost, and the choice of combiner within it follows the usual higher-criticism-versus-maximum tradeoff. The evaluation is by Monte Carlo.

\section*{Declarations}

\textbf{Ethics approval and consent to participate.} Not applicable; a methodological and simulation study with no human participants.

\textbf{Consent for publication.} Not applicable.

\textbf{Availability of data and materials.} All data are simulated or drawn from named public datasets. The reproducibility package (verification and study scripts, locked outputs, figures, the derivations companion D02, and the honest\_detection.html demonstrator) is openly archived on Zenodo as a single record; the concept DOI is to be minted at deposit.

\textbf{Competing interests.} The author develops and hosts the open-source software and associated web domains (the trialdesign.com applications) that implement related methods; no other competing interests are declared.

\textbf{Use of generative AI.} In preparing this manuscript the author used a generative-AI assistant (Claude, Anthropic) for drafting and editing prose, generating figure and simulation code, and formatting tables. All AI-assisted output was reviewed and verified by the author; every reported number regenerates deterministically from the openly deposited code, and the author takes full responsibility for the content.

\textbf{Funding.} This research received no specific grant from any funding agency in the public, commercial, or not-for-profit sectors.

\textbf{Authors' contributions.} W. J. Dwyer is the sole author and is responsible for the conception, analysis, software, and writing of this work.

\textbf{Acknowledgements.} The author thanks Rafal Wrona (independent researcher, Poland) for a careful reading of the first preprint, in particular for the counterexample that withdrew an incorrect per-cell p-value floor (correcting the scope statement of Part I and the routing threshold of Part III) and for the margin-balanced signed-rectangle construction of the sparse alternative adopted in the numerical sections.

\section*{References}

\noindent Agresti, A. (2013). Categorical Data Analysis, 3rd ed. Wiley.\par\smallskip

\noindent Anscombe, F. J. (1948). The transformation of Poisson, binomial and negative-binomial data. Biometrika 35, 246-254.\par\smallskip

\noindent Barnett, I., Mukherjee, R. and Lin, X. (2017). The generalized higher criticism for testing SNP-set effects in genetic association studies. Journal of the American Statistical Association 112, 64-76.\par\smallskip

\noindent Benjamini, Y. and Hochberg, Y. (1995). Controlling the false discovery rate: a practical and powerful approach to multiple testing. Journal of the Royal Statistical Society Series B 57(1), 289-300.\par\smallskip

\noindent Benjamini, Y. and Yekutieli, D. (2001). The control of the false discovery rate in multiple testing under dependency. Annals of Statistics 29(4), 1165-1188.\par\smallskip

\noindent Berger, R. L. and Boos, D. D. (1994). P values maximized over a confidence set for the nuisance parameter. Journal of the American Statistical Association 89, 1012-1016.\par\smallskip

\noindent Bishop, Y. M. M., Fienberg, S. E. and Holland, P. W. (1975). Discrete Multivariate Analysis: Theory and Practice. MIT Press.\par\smallskip

\noindent Chhor, J., Mukherjee, R. and Sen, S. (2024). Sparse signal detection in heteroscedastic Gaussian sequence models: sharp minimax rates. Bernoulli 30(3), 2127-2153.\par\smallskip

\noindent Darling, D. A. and Erdos, P. (1956). A limit theorem for the maximum of normalized sums of independent random variables. Duke Mathematical Journal 23, 143-155.\par\smallskip

\noindent Donoho, D. and Jin, J. (2004). Higher criticism for detecting sparse heterogeneous mixtures. Annals of Statistics 32, 962-994.\par\smallskip

\noindent Donoho, D. and Kipnis, A. (2022). Higher criticism to compare two large frequency tables, with sensitivity to possible rare and weak differences. Annals of Statistics 50(3), 1447-1472.\par\smallskip

\noindent Dwyer, W. J. (2026a). T\_root: a comprehensive closed-form statistic for independence in sparse and heterogeneous contingency tables. Manuscript under review.\par\smallskip

\noindent Dwyer, W. J. (2026b). T\_root by default: a reporting standard for independence in contingency tables. Manuscript under review.\par\smallskip

\noindent Dwyer, W. J. (2026d). Exact conditional distributions of chi-square-family statistics for two-way contingency tables, by cell-separable dynamic programming. Manuscript.\par\smallskip

\noindent Eicker, F. (1979). The asymptotic distribution of the suprema of the standardized empirical processes. Annals of Statistics 7, 116-138.\par\smallskip

\noindent Garcia-Perez, M. A. and Nunez-Anton, V. (2020). Asymptotic versus exact methods in the analysis of contingency tables: a comparison. Statistical Methods in Medical Research 29(9), 2569-2582.\par\smallskip

\noindent Goodman, L. A. (1960). On the exact variance of products. Journal of the American Statistical Association 55, 708-713.\par\smallskip

\noindent Goodman, L. A. (1962). The variance of the product of K random variables. Journal of the American Statistical Association 57, 54-60.\par\smallskip

\noindent Haberman, S. J. (1973). The analysis of residuals in cross-classified tables. Biometrics 29, 205-220.\par\smallskip

\noindent Ingster, Yu. I. (1997). Some problems of hypothesis testing leading to infinitely divisible distributions. Mathematical Methods of Statistics 6, 47-69.\par\smallskip

\noindent Jaeschke, D. (1979). The asymptotic distribution of the supremum of the standardized empirical distribution function on subintervals. Annals of Statistics 7, 108-115.\par\smallskip

\noindent Kipnis, A. (2022). Higher criticism for discriminating word-frequency tables and authorship attribution. Annals of Applied Statistics 16(2), 1236-1252.\par\smallskip

\noindent MacDonald, P. L. and Gardner, R. C. (2000). Type I error rate comparisons of post hoc procedures for I x J chi-square tables. Educational and Psychological Measurement 60, 735-754.\par\smallskip

\noindent Mukherjee, R., Mukherjee, S. and Sen, S. (2018). Detection thresholds for the beta-model on sparse graphs. Annals of Statistics 46(3), 1288-1317.\par\smallskip

\noindent Mukherjee, R., Pillai, N. S. and Lin, X. (2015). Hypothesis testing for high-dimensional sparse binary regression. Annals of Statistics 43, 352-381.\par\smallskip

\noindent Simes, R. J. (1986). An improved Bonferroni procedure for multiple tests of significance. Biometrika 73, 751-754.\par\smallskip

\noindent Wasserman, L., Ramdas, A. and Balakrishnan, S. (2020). Universal inference. Proceedings of the National Academy of Sciences 117, 16880-16890.\par\smallskip

\noindent Xie, J., Cai, T. T. and Li, H. (2011). Sample size and power analysis for sparse signal recovery in genome-wide association studies. Biometrika 98, 273-290.\par\smallskip

\section*{Data and code availability}

All verification and study scripts, the locked simulation outputs, the figures, the derivations companion (D02), and this manuscript are archived together in a single Zenodo record; the concept DOI is to be minted at deposit. Every reported number regenerates deterministically from the deposited code with the recorded seeds. Generative-AI use is disclosed per the standard M-series statement.

\end{document}